\documentclass[aps,prb,manuscript,amsmath,superscriptaddress]{revtex4}
\usepackage{graphicx}
\usepackage{color}
\usepackage{epstopdf}
\usepackage{xcolor}

\begin{document}

\title{Optical control of an individual Cr spin in a semiconductor quantum dot}

\author{L. Besombes, H. Boukari, V. Tiwari, A. Lafuente-Sampietro}
\address{Institut N\'{e}el, CNRS, Univ. Grenoble Alpes and Grenoble INP, 38000 Grenoble, France}
\author{S. Kuroda, K. Makita}
\address{University of Tsukuba, Institute of Material Science, Japan}

\begin{abstract}

Individual localized spins in semiconductors are attracting significant interest in the frame of
quantum technologies including quantum information and quantum enhanced sensing. Localized spins
of magnetic atoms incorporated in a semiconductor are particularly promising for quantum sensing.
Here we demonstrate that the spin of a Cr atom in a quantum dot (QD) can be controlled optically
and we discuss the main properties of this single spin system. The photoluminescence of
individual Cr-doped QDs and their evolution in magnetic field reveal a large magnetic anisotropy
of the Cr spin induced by local strain. This results in a splitting of the Cr spin states and in
a thermalization on the lower energy states states S$_z$=0 and S$_z$=$\pm$1. The magneto-optical
properties of Cr-doped QDs can be modelled by an effective spin Hamiltonian including the spin to
strain coupling and the influence of the QD symmetry. We also show that a single Cr spin can be
prepared by resonant optical pumping. Monitoring the intensity of the resonant fluorescence of
the QD during this process permits to probe the dynamics of the optical initialization of the
spin. Hole-Cr flip-flops induced by an interplay of the hole-Cr exchange interaction and the
coupling with acoustic phonons are the main source of relaxation that explains the efficient
resonant optical pumping. The Cr spin relaxation time is measured in the $\mu s$ range. We
evidence that a Cr spin couples to non-equilibrium acoustic phonons generated during the optical
excitation inside or near the QD). Finally we show that the energy of any spin state of an
individual Cr atom can be independently tuned by a resonant single mode laser through the optical
Stark effect. All these properties make Cr-doped QDs very promising for the development of hybrid
spin-mechanical systems where a coherent mechanical driving of an individual spin in an
oscillator is required.

\end{abstract}

\maketitle

\section{Introduction}

Individual spins in semiconductors are promising for the development of quantum technologies based
on solid state devices. Important progresses have been made recently for spins of carriers confined
in nano-structures \cite{Veldhorst2015} and for electronic and nuclear spins localized on
individual defects \cite{Schmitt2017}. Diluted magnetic semiconductor systems combining high
quality nano-structures and localized spins on transition metal elements are alternative good
candidates for the development of such single spin quantum devices. Optically active quantum dots
(QDs) containing individual or pairs of magnetic dopants can be realized both in II-VI
\cite{Besombes2004,Goryca2009,LeGall2009,LeGall2010,LeGall2011,Besombes2012} and III-V
\cite{Kudelski2007,Krebs2013} semiconductors. In these systems, since the confined carriers and
magnetic atom spins are strongly mixed, an optical excitation of the QD can affect the spin state
of the atom offering possibilities for probing and controlling of the localized spin
\cite{Govorov2005,Reiter2013}.

The variety of $3d$ transition metal magnetic elements that can be incorporated in conventional
semiconductors gives a large choice of localized electronic and nuclear spins as well as orbital
momentum \cite{Besombes2004,Kobak2014,Smolenski2016,Lafuente2016}. For a given semiconductor
nano-structure, the spin properties resulting from the exchange interaction between the confined
carriers and the incorporated magnetic dopant depend on the filling of the $3d$ orbital of the
atom. The choice of a particular magnetic element can then be adapted for a targeted application.
This approach opens a diversity of possible use of individual spins in diluted magnetic
semiconductor nano-structures for quantum information technologies or quantum sensing.

Among these magnetic atoms, chromium (Cr) is of particular interest \cite{Lafuente2016}. It usually
incorporates in intrinsic II-VI semiconductors as Cr$^{2+}$ carrying an electronic spin S=2 and an
orbital momentum L=2. Moreover, most of Cr isotopes have no nuclear spin. This simplifies the spin
level structure and the coherent dynamics of its electronic spin \cite{Vallin1974}. With bi-axial
strain, the ground state of the Cr is expected to be an orbital singlet with a spin degeneracy of
5. The non zero orbital momentum (L=2) of the Cr atom connects its electronic spin to the local
strain through the modification of the crystal field and the spin-orbit coupling. This spin to
strain coupling is expected to be more than two orders of magnitude larger than for elements
without orbital momentum like NV centers in diamond \cite{Tessier2014} or Mn atoms in II-VI
semiconductors \cite{Lafuente2015}.

In analogy with the spin structure of NV centers in diamond, the spin states S$_z=\pm1$ of a
Cr$^{2+}$ ion in a QD form a spin $qubit$ coupled to in-plane strain \cite{Lee2017}. The Cr spin is
therefore promising for the realization of hybrid spin-mechanical systems
\cite{Pigeau2015,Macquarrie2015} in which the motion of a mechanical oscillator would be coherently
coupled to the spin state of a single atom and probed or coherently controlled through this
interaction \cite{Tessier2014,Rabl2010,Ovar2014}.

This review article is organized as follows: after a short presentation of the spin properties of
magnetic atoms in semiconductors and their exchange interaction with the carriers of the host we
discuss how we can use the optical properties of a QD to probe the spin of an individual Cr. We
show that magneto-optics is an efficient tool to extract the relevant parameters of the QDs and the
strength of carriers-Cr exchange interaction. We then present optical techniques for to control the
Cr spin (resonant optical pumping and energy tuning by optical Stark effect) and discuss the
dynamics of Cr spin in the presence of optically created carriers or in the dark. We finally show
that Cr atoms in a QD forms an interesting platform for the study of the coupling of individual
spins with propagating surface acoustic phonons which are proposed as efficient quantum bus between
different kinds of qubits \cite{Schuetz2015}.

\section{Spin properties of magnetic atoms in semiconductors}

The important properties of individual magnetic atoms in diluted magnetic semiconductors arise from
(i) their fine (and hyperfine) structure, which controls the spin dynamics at zero or weak magnetic
field and (ii) the exchange interaction with the carriers of the host, which determines the
conditions of optical and electrical control. We will discuss here the origin of these two
parameters for Cr embedded in CdTe/ZnTe QDs and compare with more extensively studied Mn-doped
systems.

\subsection{Fine and hyperfine structure of a magnetic atom in a II-VI semiconductor}

\begin{figure}[hbt]
\begin{center}
\includegraphics[width=3.5in]{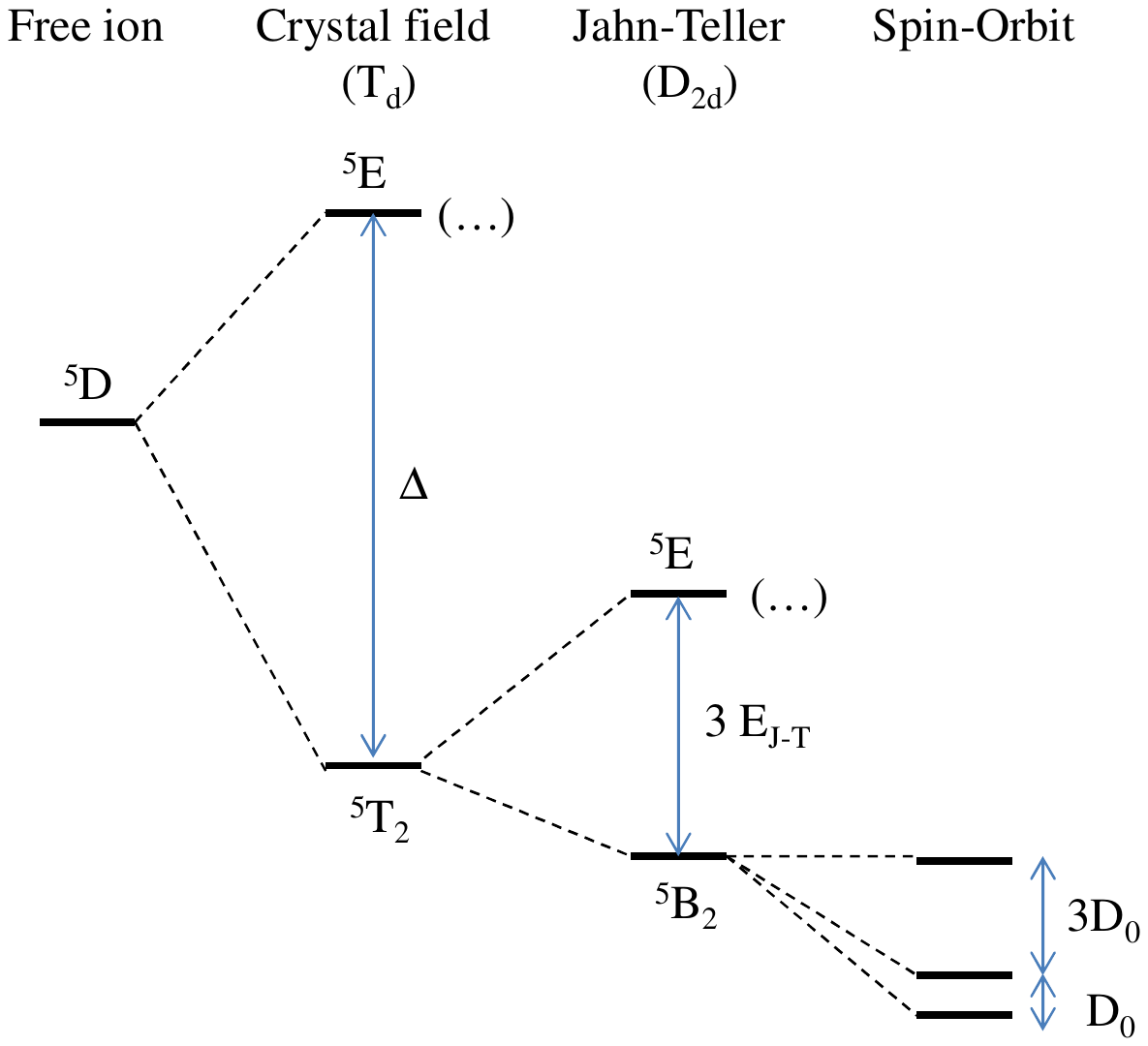}
\end{center}
\caption{Scheme of the energy level splitting of Cr$^{2+}$ at a cation site in II-VI compounds having zinc blende structure (T$_d$) with a crystal field parameter $\Delta$, a Jahn-Teller energy E$_{J-T}$ and a spin-orbit level spacing D$_0$.}
\label{FigLevelCr}
\end{figure}

Cr atoms are usually incorporated into un-doped II-VI semiconductors as Cr$^{2+}$ ions on cation sites forming a deep impurity level.
 The ground state of a free Cr$^{2+}$ is $^{5}$D with the orbital quantum number L=2 and a spin S=2 yielding a 25-fold degeneracy.
  In the crystal field of T$_{d}$ symmetry of the tetrahedral cation site in zinc-blende crystal, the degeneracy is partially lifted (see Figure~\ref{FigLevelCr}):
   the $^{5}$D term splits into 15-fold degenerate orbital triplet $^{5}$T$_{2}$ and 10-fold degenerate orbital doublet $^{5}$E.
    The Jahn-Teller distortion reduces the symmetry to D$_{2d}$ and leads to a splitting of the $^{5}$T$_{2}$ ground state into a 5-fold degenerate $^{5}$B$_{2}$ orbital singlet and a $^{5}$E orbital doublet.

The ground state orbital singlet $^{5}$B$_{2}$ is further split by the spin-orbit interaction.
 In a strain free crystal, it was found that the ground state splitting can be described by the spin effective Hamiltonian \cite{Vallin1974}:

\begin{eqnarray}
{\cal H}_{Cr,CF}={\cal D}_0S_z^2\nonumber\\
+\frac{1}{180}F[35S_z^2-30S(S+1)S_z^2+25S_z^2]\nonumber\\
+\frac{1}{6}a[S_1^4+S_2^4+S_3^4]
\label{FS}
\end{eqnarray}

\noindent with $|{\cal D}_0|\gg|a|$, $|F|$. For the studies reported here, it is a good approximation to take $a=0$ and $F=0$.
 The x, y, z principal axes were found to coincide with the cubic axes (1,2,3) giving rise to three identical sites,
 each given by (\ref{FS}) but with the z axis of each along a different cubic axis (1,2,3).
  A value of ${\cal D}_0\approx+30 \mu eV$ was estimated from Electron Paramagnetic Resonance (EPR) measurements in highly diluted bulk Cr-doped CdTe \cite{Vallin1974}.

Static biaxial compressive strain in the (001) plane, as observed in self-assembled QDs,
 reduces the symmetry to D$_{2d}$ and destabilize the Cr $3d$ orbitals $d_{xz}$ and $d_{yz}$ having an electron density pointing along the $[$001$]$ axis ($z$ axis).
  The Cr ground state is then a 5-fold spin degenerated orbital singlet formed from the $d_{xy}$ orbital. It corresponds to the Jahn-Teller ground state with a tetragonal distortion along the $[$001$]$ axis \cite{Brousseau1988}.

An additional applied stress will influence the Cr spin fine structure through the modification of the crystal field and the spin-orbit interaction.
For an arbitrary strain tensor, the general form of the Cr ground state spin effective Hamiltonian is \cite{Vallin1974}

\begin{eqnarray}
{\cal H}_{Cr,\varepsilon}=c_1e_AS_{\theta}+c_2e_{\theta}S_{\theta}+c_3e_{\epsilon}S_{\epsilon}+c_4e_{\zeta}S_{\zeta}\nonumber\\
+c_5(e_{\xi}S_{\xi}+e_{\eta}S_{\eta})
\end{eqnarray}

\noindent with S$_i$ defined as:

\begin{eqnarray}
S_{\theta}=S_{z}^2-\frac{1}{2}[S_{x}^2+S_{y}^2]\nonumber\\
S_{\epsilon}=\frac{1}{2}\sqrt{3}[S_{x}^2-S_{y}^2]\nonumber\\
S_{\xi}=S_{y}S_{z}+S_{z}S_{y}\nonumber\\
S_{\eta}=S_{x}S_{z}+S_{z}S_{x}\nonumber\\
S_{\zeta}=S_{x}S_{y}+S_{y}S_{x}
\end{eqnarray}

\noindent and $e_i$ defined similarly as:

\begin{eqnarray}
e_{\theta}=\varepsilon_{zz}-\frac{1}{2}[\varepsilon_{xx}+\varepsilon_{yy}]\nonumber\\
e_{\epsilon}=\frac{1}{2}\sqrt{3}[\varepsilon_{xx}-\varepsilon_{yy}]\nonumber\\
e_{\xi}=\varepsilon_{yz}+\varepsilon_{zy}\nonumber\\
e_{\eta}=\varepsilon_{xz}+\varepsilon_{zx}\nonumber\\
e_{\zeta}=\varepsilon_{xy}+\varepsilon_{yx}\nonumber\\
e_A=\varepsilon_{xx}+\varepsilon_{yy}+\varepsilon_{zz}
\end{eqnarray}

\noindent where the $\epsilon_{ij}$ are components of the strain tensor. For a flat CdTe
self-assembled QDs in ZnTe with a dominant in-plane biaxial strain we have a strain tensor:

\begin{equation}\label{H-exc}
\mathcal{\varepsilon}_{ij} = \left(
\begin{array}{ccc}
\varepsilon_{\parallel}          &0                              &0                \\
0                                &\varepsilon_{\parallel}        &0                \\
0                                &0                              &\varepsilon_{zz} \\
\end{array}\right)
\end{equation}

\noindent with $\varepsilon_{zz}=-2\frac{C_{12}}{C_{11}}\varepsilon_{\parallel}$ and C$_{ij}$ the
stiffness constants for CdTe.

\begin{table}[htb] \centering
\caption{Values for spin to strain coupling coefficients of Cr in bulk CdTe (in $meV$) extracted from EPR measurements in ref. \cite{Vallin1974}.}
\label{table1}\renewcommand{\arraystretch}{1.0}
\begin{tabular}{p{1.6cm}p{1.6cm}p{1.6cm}p{1.6cm}p{1.6cm}}
\hline\hline
c$_{1}$ & c$_{2}$ & c$_{3}$  & c$_{4}$  & c$_{5}$ \\
-0.25$\pm$2 & +4.9 $\pm$2& -1.2$\pm$0.5 & +4.9$\pm$2 & +3.7$\pm$1.25 \\
\hline\hline
\end{tabular}
\end{table}

For this strain configuration, the Cr fine structure is controlled by the spin-lattice coupling
coefficients c$_1$ (symmetric coefficient) and c$_2$ (tetragonal coefficients). The strain-coupling
coefficients estimated from EPR measurements in bulk Cr doped CdTe are listed in table
\ref{table1}. The strain controlled part of the spin Hamiltonian ${\cal H}_{\varepsilon}$ becomes:

\begin{eqnarray}
{\cal H}_{Cr,\varepsilon\parallel}= \frac{3}{2}\varepsilon_{\parallel}[2c_1(1-\frac{C_{12}}{C_{11}})-c_2(1+2\frac{C_{12}}{C_{11}})]S_z^2=D_0S_z^2
\end{eqnarray}

\noindent where we can estimate a magnetic anisotropy D$_0\approx$ 1$\pm$0.6 meV from the values of the spin to strain coupling coefficients in pure
CdTe (table \ref{table1}) and $\varepsilon_{\parallel}=(a_{ZnTe}-a_{CdTe})/a_{CdTe}\approx-5.8\%$.

An anisotropy of the strain in the QD plane (001) would affect the Cr fine structure through the tetragonal coefficients c$_3$ and c$_4$. The largest
spin to strain coupling is expected for a strain along $[$110$]$ direction (coefficient $c_4$) with $\varepsilon_{xy}=\varepsilon_{yx}\neq0$ and
$\varepsilon_{xx}=\varepsilon_{yy}=0$. This interaction can be described by an additional term in the spin-strain Hamiltonian

\begin{eqnarray}
{\cal H}_{Cr,\varepsilon\perp}=c_4(\varepsilon_{xy}+\varepsilon_{yx})(S_xS_y+S_yS_x)=c_4(\varepsilon_{xy}+\varepsilon_{yx})\frac{1}{2i}(S_+^2-S_-^2)
\label{epsilonperp}
\end{eqnarray}

This anisotropy term couples spin states separated by two units and in particular S$_z$=+1 and S$_z$=-1 which are initially degenerated in the
absence of magnetic field. As we will see in the last section, such term, together with terms proportional to $\epsilon_{xx}-\epsilon_{yy}$, can be
exploited to induce a strain mediated coherent coupling between a mechanical oscillator and a Cr spin.

The situation is different for a Mn atom which, incorporated as a Mn$^{2+}$ ion in II-VI compounds,
has no orbital momentum. As the ground state of the  Mn$^{2+}$ has no orbital degeneracy it is not
affected by the Td crystal field nor by the reduction of its symmetry by biaxial strain. However
the spin degeneracy is lifted by a combination of spin-orbit interaction and reduced symmetry of
the crystal field. As in the case of Cr this results in a splitting of the spin levels according to
$D_0S_z^2$ but with $D_0$ at least two orders of magnitude weaker than in the Cr case (values
around 7$\mu eV$ were measured in CdTe/ZnTe QDs \cite{Goryca2014,Jamet2013}. All the Mn stable
isotopes carries a nuclear spin I=5/2. This nuclear spin couples to the $3d$ electrons via the
hyperfine interaction ${\cal A}I.S$ with ${\cal A}\approx0.7\mu eV$. The fine and hyperfine
splitting have similar values and this results in a complex 36 spin level structure responsible of
the rich spin dynamics observed for an individual Mn atom in a QD \cite{Goryca2014,Jamet2013}.

\subsection{Exchange interaction between carriers and a magnetic atom}

Diluted magnetic semiconductors are characterised by a large exchange interaction between the
localized spins of the magnetic atoms and the carriers of the host. The electron-magnetic atom
coupling for carriers at the center of the Brillouin zone arises from the standard short range
exchange interaction. It is then ferromagnetic and a value of the exchange energy
N$_0\alpha\approx0.2eV$ (where $N_0$ is the number of unit cell in a normalized volume) is found
for most of the transition metals incorporated in wide band gap II-VI compounds \cite{Kacman2001}.

The exchange interaction with the hole spin is usually larger than the interaction with the
electron spin and arises from two mechanisms: (i) a ferromagnetic coupling resulting from the short
range exchange interaction and (ii) a spin dependent hybridization of the $d$ orbital of the
magnetic atom and the $p$ orbital of the host semiconductor, the so-called kinetic exchange.

The $p-d$ hybridization is strongly sensitive to the energy splitting between the $3d$ levels of
the atom and the top of the valence band. This hybridization of course significantly depends on the
considered transition metal element (i.e. filling of the $3d$ orbital) and, for a given magnetic
element, on the semiconductor host \cite{Kossut}. It can be either ferromagnetic or
anti-ferromagnetic depending on the relative position of the $d$ levels and the top of the valence.

In the case of Mn, the Mn$^{2+/3+}$ donor level is located far within the valence band and the
kinetic exchange results in an anti-ferromagnetic hole-Mn coupling. The p-d hybridization has the
main contribution to the hole-Mn exchange and the overall interaction is anti-ferromagnetic. A
value of N$_0\beta\approx-0.88eV$ has been measured for Mn in CdTe \cite{Furdyna1988}.

For a Cr atom in the 3$d^4$ configuration, it has been demonstrated \cite{Blinowski1992,Kacman2001}
that the exchange interaction between a Cr atom with a Jahn-Teller distortion oriented along the
[001] axis (z-axis) and a heavy-hole with a $z$ component of its total angular momentum
J$_z$=$\pm$3/2 can be expressed in the form:

\begin{eqnarray}
H_{ex}=-\frac{1}{3}S_zJ_zB_4
\label{exchange}
\end{eqnarray}

\noindent where $B_N$ (with N=4 for Cr) is given by

\begin{eqnarray}
B_N=-\frac{V_{pd}^2}{S}[\frac{1}{\varepsilon_p+E_{N-1}^{S-1/2}-E_{N}^{S}}+\frac{1}{E_{N+1}^{S-1/2}-E_{N}^{S}-\varepsilon_p}]
\label{BN}
\end{eqnarray}

\noindent with $E_N^S$ the unperturbed energy of the $d$ shell with N electrons and total spin S, V$_{pd}$ the hybridization constant between the $d$
orbital of the impurity and the $p$ orbital of the semiconductor host and $\varepsilon_p$ the energy of the top of the valence band
\cite{Kacman2001}. This Cr site orientation along the [001] axis corresponds to the CdTe/ZnTe QDs where a Cr atom can be optically detected. Such
dots are characterized by a large biaxial strain which dominates the Jahn-Teller effect and orients the Cr spin along the QD growth axis
\cite{Brousseau1988}.

In the expression (\ref{BN}) controlling the amplitude of the hole-Cr exchange, the first denominator corresponds to the energy $e_1$ required to
transfer an electron from the $d$ shell of the Cr atom to the valence band reducing the total spin from S to S-1/2 (and reducing the number of
electrons in the $d$ shell from N to N-1). The denominator of the second term is the energy $e_2$ required to transfer an electron from the top of
the valence band to the $d$ shell also with a reduction by 1/2 of the total spin (and an increase of the number of electrons in the $d$ shell from N
to N+1). This last energy $e_2$ includes the electron-electron exchange interaction in the $d$ shell (at the origin of Hund rule) and is then large
(a few eV) and always positive.

The sign of $B_4$ controls the sign of the hole-Cr kinetic exchange interaction: a negative $B_4$
corresponds to an anti-ferromagnetic interaction whereas a positive $B_4$ will give rise to a
ferromagnetic interaction. If e$_1$ is negative ($E_{N}^{S}-E_{N-1}^{S-1/2}>\varepsilon_p$) and
$1/e_1<-1/e_2$, $B_4$ is positive. The donor transition Cr$^{2+}$ to Cr$^{3+}$ is within the band
gap of the semiconductor and the hole-Cr exchange interaction is ferromagnetic. This is the
situation reported until now for all the studied bulk II-VI compounds containing diluted Cr atoms
\cite{Blinowski1996}. In particular, Cr-doped bulk ZnTe exhibits a large ferromagnetic exchange
interaction. This is consistent with the optical observation of the Cr$^{2+/3+}$ donor level about
0.2 eV above the top of the valence band in ZnTe \cite{Kuroda2007}.

For a positive value of $e_1$ ($E_{N}^{S}-E_{N-1}^{S-1/2}<\varepsilon_p$), $B_4$ is negative and
the hole-Cr exchange interaction is anti-ferromagnetic. This corresponds to a donor level
Cr$^{2+/3+}$ within the valence band. According to equation (\ref{BN}), a slight change of the
value of e$_1$ around 0 can abruptly change the hole-Cr exchange from a large ferromagnetic to a
large anti-ferromagnetic value. However, one should note that the calculations leading to the
expression of the kinetic exchange interaction (\ref{exchange}) contain two important
approximations. First, the influence of the crystal field and strain modified crystal field on the
magnetic atom $d$ orbital is neglected. Secondly, the model is based on a perturbation approach
which is not particularly well adapted when $e_1$ is close to zero as expected for Cr in CdTe or
ZnTe.

The hole-Cr exchange interaction has never been measured in bulk CdTe. However, the energy level of
a transition-metal impurity does not significantly change between materials with common anion
\cite{Kossut}. The valence band offset between ZnTe and CdTe being around 0.1 eV
\cite{Continenza1994} one could also expect that the Cr donor transition in bulk CdTe could be very
close but slightly above the top of the valence band. This should give rise to a ferromagnetic
hole-Cr exchange interaction as observed for Cr-doped ZnTe. However, whereas the acceptor level
Cr$^{2+/1+}$ has clearly been optically identified in bulk CdTe, the donor level Cr$^{2+/3+}$ has
never been observed. This suggests, following reference \cite{Rzepka1993}, that it could be
resonant with the valence band. The resulting hole-Cr exchange interaction would then be large and
anti-ferromagnetic.

\section{Optical probing of the spin state of an individual Cr atom}

To optically access individual magnetic atoms, Cr are randomly introduced in CdTe/ZnTe
self-assembled QDs grown by Molecular Beam Epitaxy on ZnTe (001) substrates following the procedure
described in ref.\cite{Wojnar2011}. The amount of Cr is adjusted to optimize the probability to
detect QDs containing 1 or a few Cr atoms. The emission of individual QDs, induced by optical
excitation with a dye laser tuned on resonance with an excited state of the dots
\cite{Besombes2014}, is studied in magnetic fields (up to 11 T) by optical micro-spectroscopy in
Faraday configuration.

\subsection{Exciton-Cr in a CdTe/ZnTe quantum dot}

\begin{figure}[hbt]
\begin{center}
\includegraphics[width=4.5in]{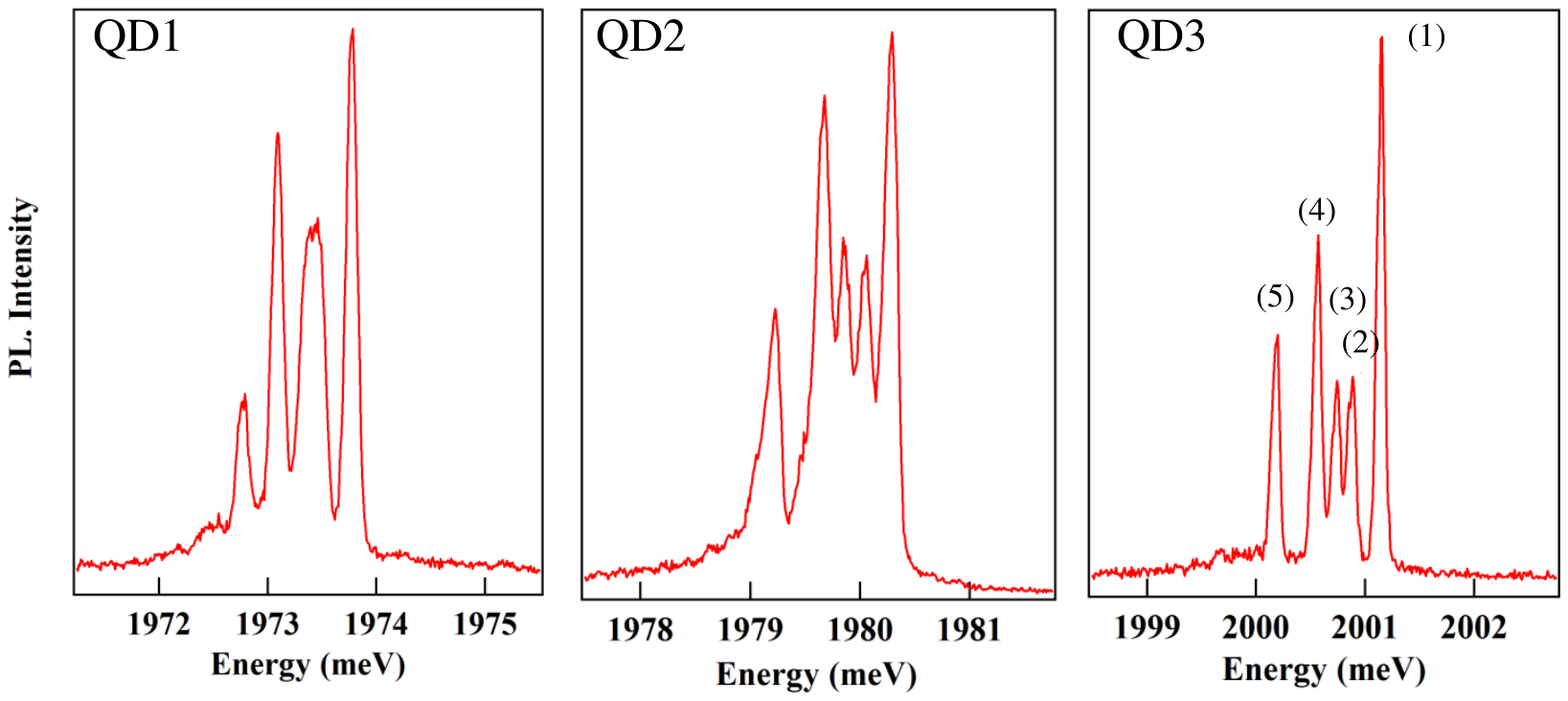}
\end{center}
\caption{PL of exciton-Cr at low temperature (T=5K) in three different QDs, QD1, QD2 and QD3.}
\label{FigQDCr}
\end{figure}

The low temperature (T=5K) PL of the neutral exciton coupled to a single Cr spin (X-Cr) of three
individual Cr-doped QDs (QD1, QD2 and QD3) are reported in Figure~\ref{FigQDCr}. Three main
emission lines are observed for X-Cr. The relative intensities of the lines and their splitting
change from dot to dot. For some of the dots, a splitting of the central line is observed and an
additional line appears on the low energy side of the X-Cr spectra. All these features result from
the exchange coupling of the electron and hole spins with a single Cr spin.

When an electron-hole (e-h) pair is injected in a Cr-doped QD, the bright excitons are split by the
exchange interaction between the spins of Cr and carriers. In flat self-assembled QDs, the
heavy-holes and light-holes are separated in energy by the biaxial strain and the confinement. In a
first approximation, the ground state in such QD is a pure heavy-hole (J$_z$=$\pm$3/2) exciton and
the exchange interaction with the Cr spin S is described by the spin Hamiltonian

\begin{eqnarray}
{\cal H}_{c-Cr}=I_{eCr}\vec{S}\cdot\vec{\sigma}+I_{hCr}S_zJ_z
\end{eqnarray}

\noindent with $\vec{\sigma}$ the electron spin and J$_z$ the hole spin operator. I$_{eCr}$ and I$_{hCr}$ are, respectively, the exchange integrals
of the electron and the hole spins with the Cr spin. These exchange energies depend on the exchange constant of the $3d$ electrons of the Cr with the
carriers in CdTe and on the overlap of the Cr atom with the confined carriers.

\begin{figure}[hbt]
\begin{center}
\includegraphics[width=2.5in]{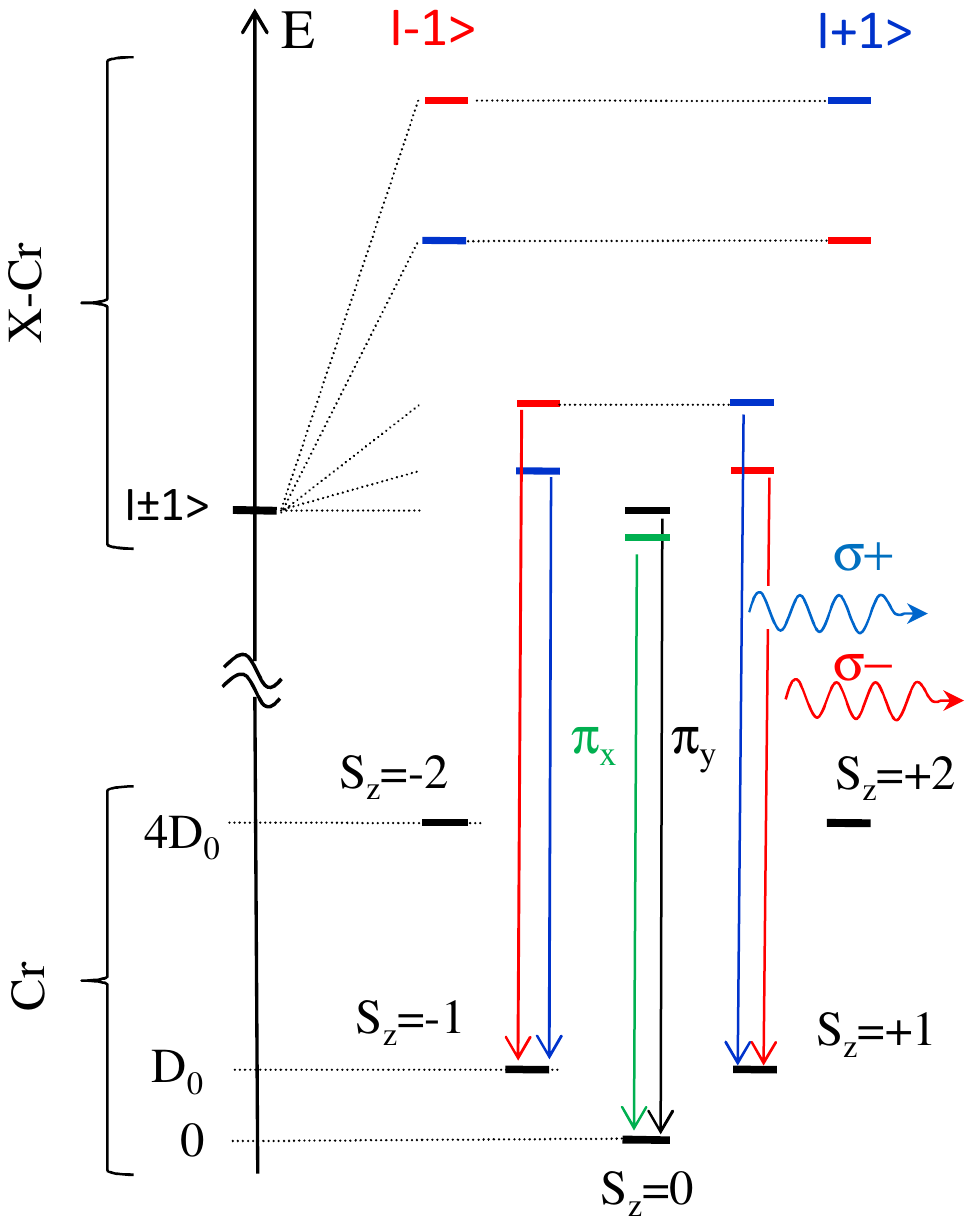}
\end{center}
\caption{Diagram of the energy levels of the ground state (Cr alone) and the bright exciton state coupled to the spin of the Cr (X-Cr) in a strained Cr-doped QD.
Vertical arrows indicates the main optical transition with their dominant polarization either linear ($\pi$) or circular ($\sigma$).}
\label{Figniveaux}
\end{figure}

For highly strained CdTe/ZnTe QDs with a weak hole confinement, the strain induced energy splitting
of the Cr spin $D_0S^2_z$ is much larger than the exchange energy with the confined carriers
($D_0\gg |I_{hCr}|>|I_{eCr}|$). The exchange interaction with the exciton acts as an effective
magnetic field which further splits the Cr spins states S$_z$=$\pm$1 and S$_z$=$\pm$2. The
resulting X-Cr energy levels are presented in Figure~\ref{Figniveaux}. The exciton recombination
does not affect the Cr atom and its spin is conserved during the optical transitions. Consequently,
the large strained induced splitting of the Cr spin is not directly observed in the optical
spectra. However, at low temperature, the Cr spin thermalize on the low energy states S$_z$=0 and
S$_z$=$\pm$1. This leads to a PL dominated by three contributions: A central line corresponding to
S$_z$=0 and the two outer lines associated with S$_z$=$\pm$1 split by the exchange interaction with
the carriers.

\begin{figure}[hbt]
\begin{center}
\includegraphics[width=3.5in]{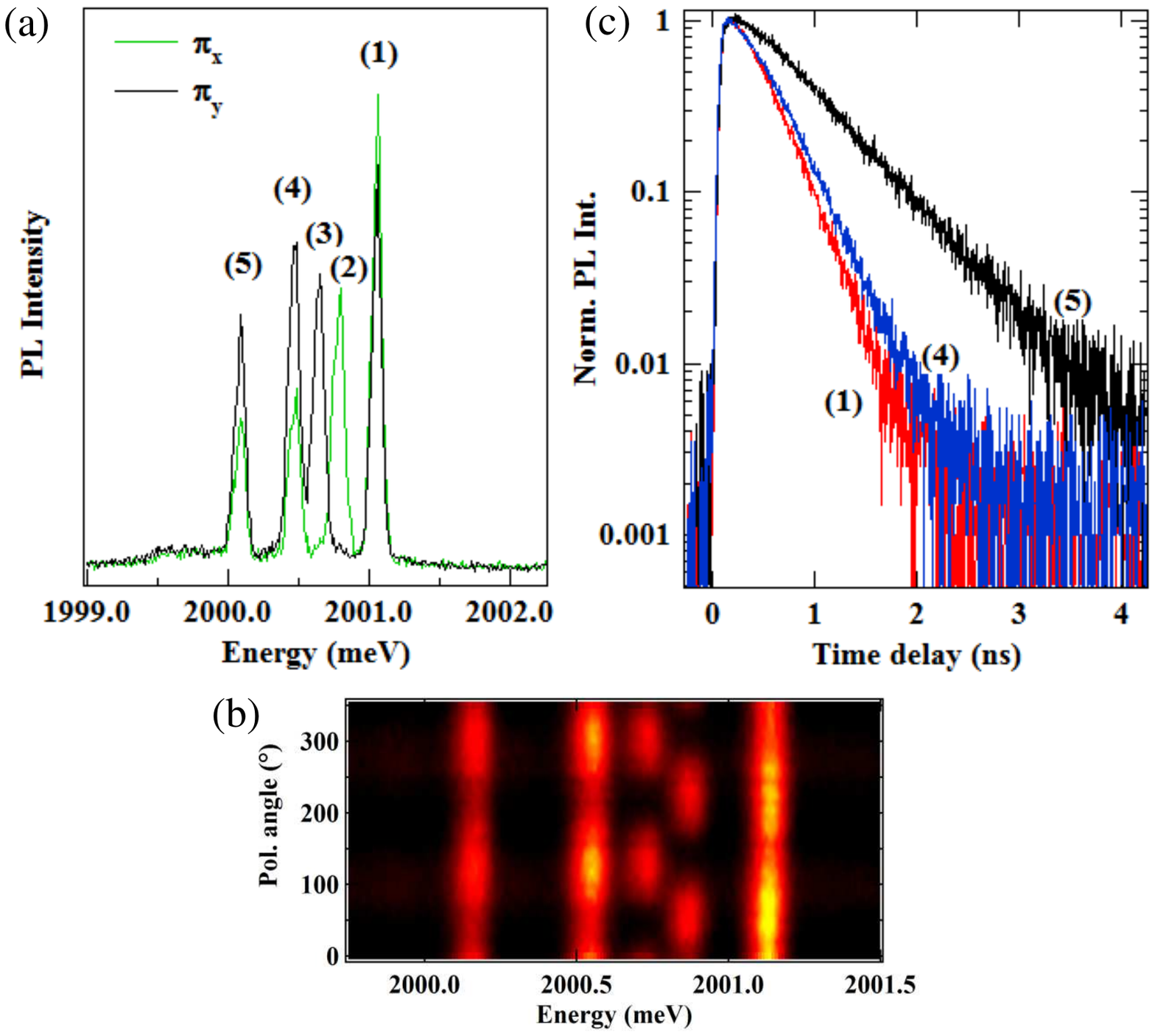}
\end{center}
\caption{(a) Low temperature PL spectra of QD3 recorded in linear polarization along two orthogonal directions. (b) Linear polarization PL intensity map of QD3.
The 0 degree polarization direction corresponds to a cleaved edge of the sample ([110] or [1-10] crystallographic directions).
(c) Time-resolved PL of QD3 recorded on the two outer lines (line (1) and (4)) and on the low energy line (5).}
\label{FigPolLine}
\end{figure}

As presented in the case of QD3 in Figure\ref{FigPolLine}, most of the Cr-doped QDs exhibit a
linear polarization dependence. The central line is split and linearly polarized along two
orthogonal directions (lines (2) and (3)). As in non-magnetic QDs, this results from a coupling of
the two bright excitons by (i) the long range e-h exchange interaction in QDs with an in-plane
shape anisotropy and/or (ii) the short range exchange interaction in the presence of valence band
mixing. This anisotropic e-h exchange interaction energy mixes the bright exciton associated with
the same Cr spin state inducing an additional splitting between them. The mixing is maximum for the
central bright exciton lines which are initially degenerated. The outer lines (lines (1) and (4))
are also partially linearly polarized but the influence of the e-h exchange interaction is here
attenuated by the initial splitting of the bright excitons induced by the exchange interaction with
the Cr spin $S_z=\pm1$.

In many Cr-doped QDs, an additional line appears on the low energy side of the PL spectra at zero
magnetic field. As presented in Figure~\ref{FigPolLine} for QD3, this line presents a longer
lifetime. It arises from a dark exciton (total angular momentum $\pm2$) which acquires some
oscillator strength by a mixing with a bright exciton interacting with the same Cr spin state
\cite{Lafuente2016}. This dark/bright exciton mixing is induced by the e-h exchange interaction in
a confining potential of low symmetry \cite{Zielinski2015}.

Let us note finally that in the presence of electrical doping and/or optical excitation, charge transfer from the Cr atom to the band of the
semiconductor or to other localized levels may occur leading to changes in the $d$ shell configuration and in the charge states of the Cr. This is in
particular the case for Cr in ZnTe where the donor and level Cr$^{2+/3+}$ and the acceptor level Cr$^{2+/1+}$ are both within the band gap. Such
fluctuation of the charge state of a Cr atom has been observed optically for Cr atoms located in the ZnTe barriers close to a QD \cite{Besombes2019}.
Fluctuation of the charge of the Cr between Cr$^{2+}$ and Cr$^{+}$ could also be possible for a Cr in CdTe as the acceptor level Cr$^{2+/1+}$ is
within the band gap close to the conduction band. Such fluctuation between a $3d^4$ (S=2) and a $3d^5$ (S=5/2) magnetic atom has never been
identified until now in the spectra of individual QDs.

\subsection{Magneto optical properties of Cr-doped quantum dots}

\begin{figure}[hbt]
\begin{center}
\includegraphics[width=4.5in]{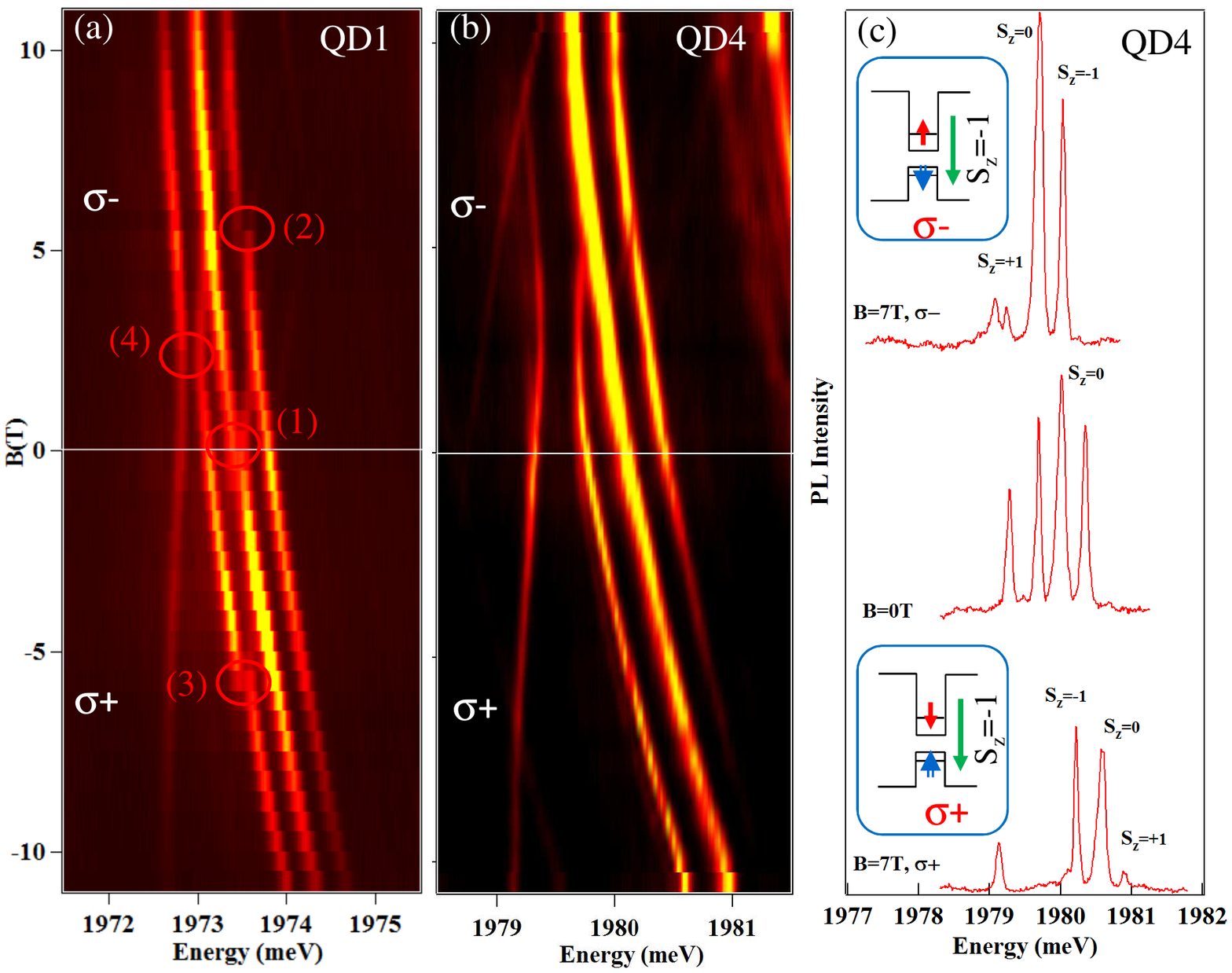}
\end{center}
\caption{PL intensity map of the evolution under a longitudinal magnetic field of the circularly polarized PL of X-Cr in QD1 (a) and QD4 (b).
(c) Circularly polarized PL spectra of QD4 at B$_z$=0T and B$_z$=7T in $\sigma+$ and $\sigma-$ polarizations. The insets present the spin configurations for the most intense PL lines.}
\label{FigBtherm}
\end{figure}

The structure of the energy levels in QDs containing a Cr$^{2+}$ ion is confirmed by the evolution
of the PL spectra in magnetic field. The circularly polarized PL of QD1 and QD4 under a magnetic
field applied along the QD growth axis are presented in Figure~\ref{FigBtherm}. Under a magnetic
field the exciton states $|\Downarrow_h\uparrow_e\rangle=|-1\rangle$ and
$|\Uparrow_h\downarrow_e\rangle=|+1\rangle$ are split by the Zeeman energy. This splitting can
compensate the exciton splitting induced by the exchange interaction with the Cr (anti-crossing (1)
at B$_z$=0 T in Figure~\ref{FigBtherm}(a)) \cite{Leger2005}. For QD1, this results in an
anti-crossing of $|+1\rangle$ and $|-1\rangle$ excitons due to the e-h exchange interaction around
B$_z$=6 T observed both in $\sigma$+ and $\sigma$- polarizations (anti-crossing (2) and (3) in
Figure~\ref{FigBtherm}(a)).

\begin{figure}[hbt]
\begin{center}
\includegraphics[width=2.5in]{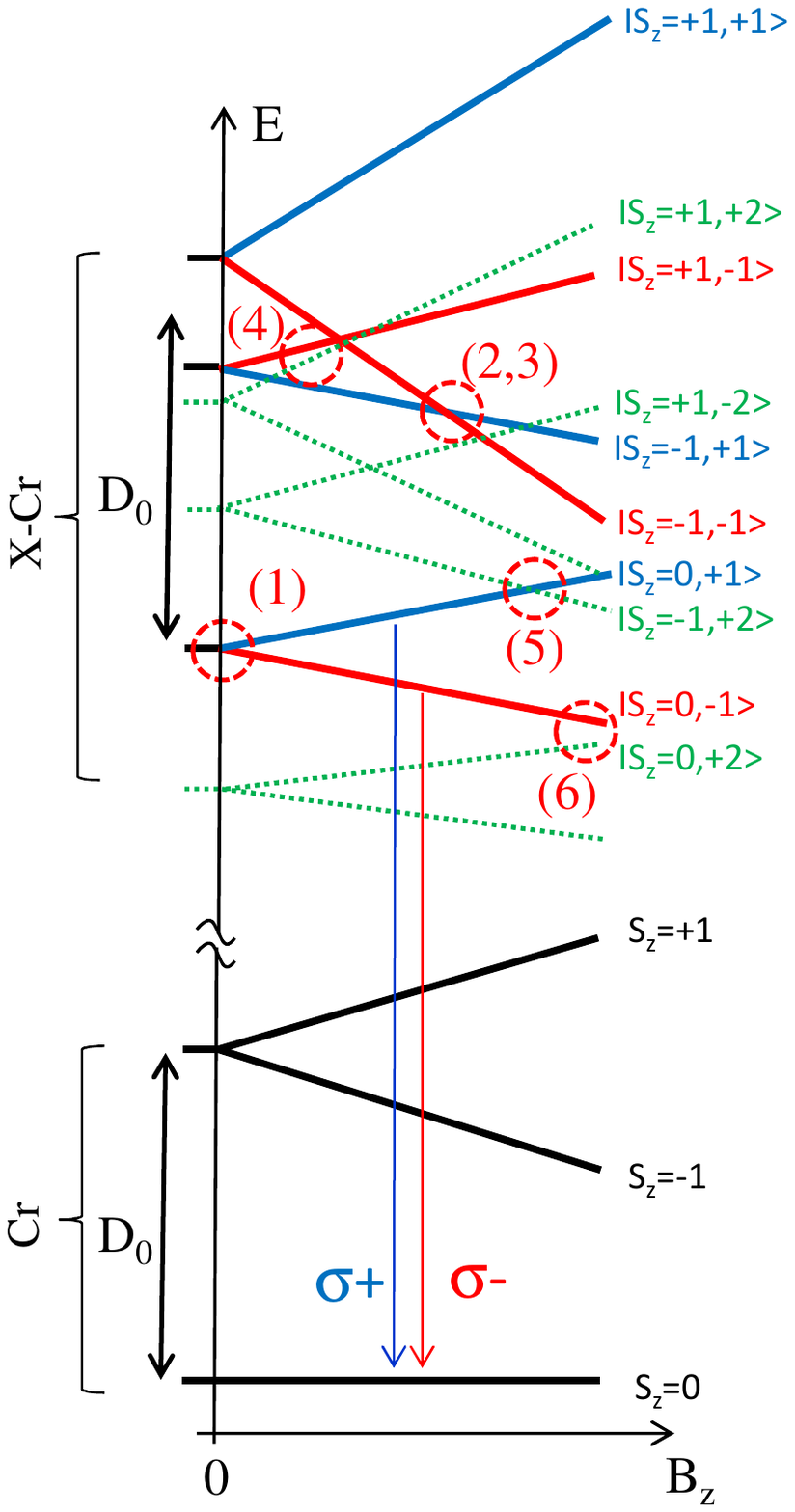}
\end{center}
\caption{Diagram of the energy levels of the Cr alone (Cr) and of an exciton-Cr (X-Cr) in a Cr-doped QD under a longitudinal magnetic field.}
\label{FigniveauxB}
\end{figure}

An anti-crossing of the low energy dark exciton with the bright excitons is observed under B$_z$ in
$\sigma$- polarization (anti-crossing (4) in Figure~\ref{FigBtherm}(a)). As illustrated in
Figure~\ref{FigniveauxB} this anti-crossing arises from a mixing of the bright and dark excitons
interacting with the same Cr spin state. Observed in $\sigma$- polarization, it corresponds to the
mixing of the exciton states $|-1\rangle$ and $|+2\rangle$ coupled to the Cr spin S$_z$=-1. This
dark/bright exciton coupling $\delta_{12}$ is induced by the e-h exchange interaction in a
confining potential of reduced symmetry (lower than C$_{2v}$) \cite{Zielinski2015}. In such
symmetry, the dark excitons acquire an in-plane dipole moment which lead to possible optical
recombination at zero magnetic field \cite{Bayer2002} as observed in QD3 and QD4. The oscillator
strength of this "dark exciton" increases as the initial splitting between $|-1\rangle$ and
$|+2\rangle$ excitons is reduced by the magnetic field (Figure\ref{FigniveauxB}).

At moderate excitation power, a variation under magnetic field of the intensity distribution
between the high and low energy lines can be observed in some of the Cr-doped QDs (See QD4 in
Figure~\ref{FigBtherm}). Under a magnetic field applied along the QD growth axis, a maximum of PL
intensity is observed on the high energy side of the exciton-Cr spectra in $\sigma-$ polarization
and on the low energy side in $\sigma+$ polarization. This change in the intensity distribution is
more or less pronounced from one dot to another (see QD1 and QD4 in Figure~\ref{FigBtherm}) but
always presents the same tendency and easier to observe at low excitation intensity
\cite{Lafuente2018}. Such distribution, with a maximum of PL intensity on the high energy side in
$\sigma$- polarization (which shifts to low energy under B$_z>$0) and on the low energy side in
$\sigma$+ polarization, is identical to the one observed in CdTe/ZnTe QDs doped with a single Mn
atom. In these systems, the hole-Mn exchange interaction is known to be anti-ferromagnetic
\cite{Besombes2005} and approximately four times larger than the ferromagnetic electron-Mn exchange
interaction.

Under a longitudinal magnetic field the Cr spin states S$_z=\pm1$ in the empty QD are split by the Zeeman energy $g_{Cr}\mu_BB_z$ with a Lande factor
$g_{Cr}\approx2$. Among the Zeeman doublet S$_z=\pm1$, the Cr spin thermalizes on the lowest energy Cr spin state $S_z=-1$ at low temperature. At
moderate excitation power where the exciton-Cr PL intensity distribution is not dominated by the carriers-Cr spin-flips, this population distribution
can be partially mapped on the exciton-Cr PL. When an unpolarized exciton is injected in the QD, the energy position of the most populated states
with S$_z=-1$, on the high energy or the low energy side of the exciton PL spectra, depends on the sign of the exciton-Cr exchange interaction. This
exchange interaction results from the sum of the electron-Cr and hole-Cr exchange interactions.

The intensity distribution observed under magnetic field in the studied Cr-doped QDs (see QD4 in
Figure\ref{FigBtherm}) shows that the exciton-Cr state
$|S_z=-1\rangle|\Downarrow_h\uparrow_e\rangle$ is at high energy whereas the state
$|S_z=-1\rangle|\Uparrow_h\downarrow_e\rangle$ is at low energy. This corresponds, for a CdTe/ZnTe
QD, to an anti-ferromagnetic exchange interaction between the heavy-hole and the Cr spins. The sign
of this interaction could however be different in bulk CdTe. For a CdTe QD in a ZnTe barrier, the
biaxial strain can decrease the energy of the ground heavy hole levels, $e_1$ can become negative
and the resulting exchange interaction can therefore be anti-ferromagnetic. The strain induced
modification of the crystal field can also significantly influence the energy level of the $d$
orbital of the Cr and modify the kinetic exchange with the nearby hole spins at the top of the
valence band. A more detailed model should be developed to properly describe the influence of the
hybridisation in these strained and confined systems where the $d$ levels of the magnetic atom are
close in energy with the edge of the valence band.

\begin{figure}[hbt]
\begin{center}
\includegraphics[width=3.0in]{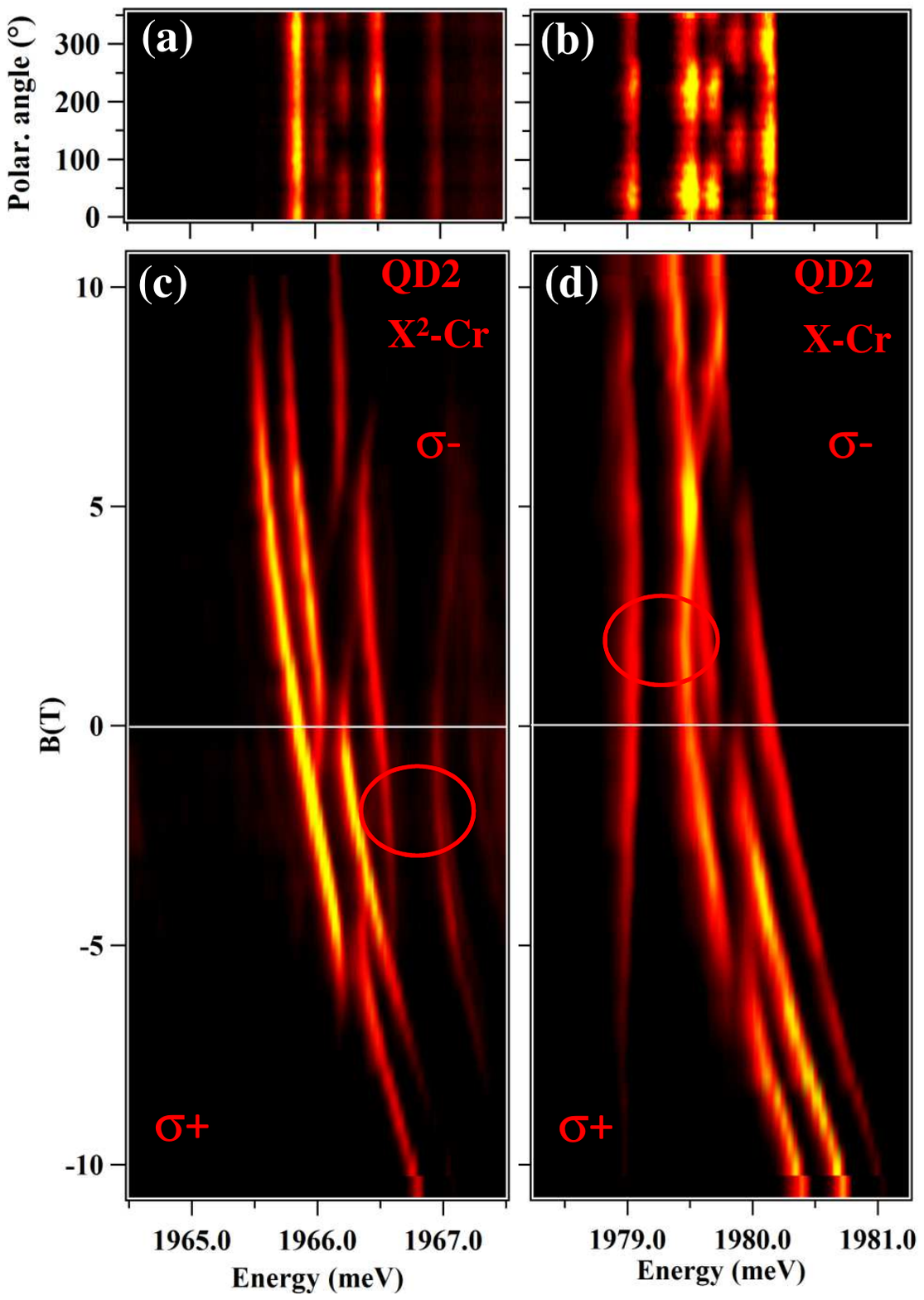}
\end{center}
\caption{Linear polarization intensity map of X$^2$-Cr (a) and X-Cr (b) in QD2. (c) and (d): intensity map of the longitudinal magnetic field dependence of X$^2$-Cr and X-Cr respectively.}
\label{FigXX2}
\end{figure}

Investigating both the biexciton ($X^2$) and the exciton ($X$) in the same Cr-doped QD, we can also
analyze the impact of the carrier-Cr interaction on the fine structure of the Cr spin. The magnetic
field dependence of X$^2$-Cr and X-Cr emissions in QD2 are presented as a contour plot in
Figure~\ref{FigXX2}(c) and (d) respectively. The PL under magnetic field of X-Cr and X$^2$-Cr
present a mirror symmetry. In particular, the dark/bright exciton mixing observed around B$_z$=2.5T
on the low energy side of the PL in $\sigma-$ polarization for X-Cr is observed on the high energy
side in $\sigma+$ polarization for X$^2$-Cr (circles in Figure~\ref{FigXX2}(c) and (d)).

\subsection{Modeling of Cr-doped quantum dots}

To estimate the relevant parameters of a the Cr-doped QDs from the optical spectra the, we
calculated the magneto-optic behavior of Cr-doped QDs by diagonalizing the complete Hamiltonian of
the electron-hole-Cr system. We consider the general case of QDs with a symmetry lower than
C$_{2v}$ (truncated ellipsoidal lens for instance \cite{Zielinski2015}), and take into account the
influence of this reduced symmetry on the valence band and on the e-h exchange interaction.

The complete electron-hole-Cr Hamiltonian in a QD (${\cal H}_{X-Cr}$) can be separated into six
parts:

\begin{eqnarray}
\label{X-Cr} {\cal H}_{X-Cr}={\cal H}_{Cr,\varepsilon}+{\cal H}_{c-Cr}+{\cal H}_{mag}+{\cal H}_{e-h}+{\cal H}_{band}+{\cal H}_{scat}
\end{eqnarray}

${\cal H}_{Cr,\varepsilon}$ describes the fine structure of the Cr atom and its dependence on local strain as presented in section 2.

${\cal H}_{c-Cr}$ describes the coupling of the electron and hole with the Cr spin. It reads

\begin{eqnarray}
\label{c-Cr} {\cal H}_{c-Cr}= I_{eCr}\overrightarrow{S}.\overrightarrow{\sigma}+I_{hCr}\overrightarrow{S}.\overrightarrow{J}
\end{eqnarray}

\noindent with $I_{eCr}$ and $I_{hCr}$ the exchange integrals of the electron ($\overrightarrow{\sigma}$) and hole ($\overrightarrow{J}$) spins with
the Cr spin ($\overrightarrow{S}$).

An external magnetic field couples via the standard Zeeman terms to both the Cr spin and carriers spins and a diamagnetic shift of the electron-hole
pair can also be included resulting in

\begin{eqnarray}
\label{cmag3} {\cal H}_{mag}=g_{Cr}\mu_B\overrightarrow{B}.\overrightarrow{S}+g_{e}\mu_B\overrightarrow{B}.\overrightarrow{\sigma}+g_{h}\mu_B\overrightarrow{B}.\overrightarrow{J}+\gamma B^2
\end{eqnarray}

The electron-hole exchange interaction, ${\cal H}_{e-h}$, contains the short range and the long
range parts. The short range contribution is a contact interaction which induces a splitting
$\delta_0^{sr}$ of the bright and dark excitons and, in the reduced symmetry of a zinc-blend
crystal ($T_d$), a coupling  $\delta_2^{sr}$ of the two dark excitons. The long range part also
contributes to the bright-dark splitting by an energy $\delta_0^{lr}$. In QDs with C$_{2v}$
symmetry (ellipsoidal flat lenses for instance \cite{Zielinski2015}) the long range part also
induces a coupling $\delta_1$ between the bright excitons. Realistic self-assembled QDs have
symmetries which can deviate quite substantially from the idealized shapes of circular or
ellipsoidal lenses. For a $C_{s}$ symmetry (truncated ellipsoidal lens), additional terms coupling
the dark and the bright excitons have to be included in the electron-hole exchange Hamiltonian.
Following Ref. \cite{Zielinski2015}, the general form of the electron-hole exchange Hamiltonian in
the heavy-hole exciton basis $|+1\rangle$, $|-1\rangle$, $|+2\rangle$, $|-2\rangle$ for a low
symmetry QD (C$_s$) is

\begin{eqnarray}
\label{Heh}
\mathcal{H}_{e-h}=
\frac{1}{2} \left(
\begin{array}{cccc}
-\delta_0                               &e^{i\pi/2}\delta_1              &e^{i\pi/4}\delta_{11}        &-e^{i\pi/4}\delta_{12}\\
e^{-i\pi/2}\delta_1                     &-\delta_0                       &e^{-i\pi/4}\delta_{12}       &-e^{-i\pi/4}\delta_{11}\\
e^{-i\pi/4}\delta_{11}                  &e^{i\pi/4}\delta_{12}           &\delta_0                     &\delta_2\\
-e^{-i\pi/4}\delta_{12}                 &-e^{i\pi/4}\delta_{11}          &\delta_2                     &\delta_0\\
\end{array}\right)
\end{eqnarray}

The terms $\delta_{11}$ and $\delta_{12}$, not present in symmetry C$_{2v}$, give an in-plane
dipole moment to the dark excitons \cite{Bayer2002}. The term $\delta_{12}$ which couples
$|\pm1\rangle$ and $|\mp2\rangle$ excitons respectively is responsible for the dark-bright
anti-crossing observed on the low energy side (lines (4) and (5)) of the emission of Cr-doped QDs.

The band Hamiltonian, ${\cal H}_{band}=E_g+\mathcal{H}_{vbm}$, stands for the energy of the electrons (i.e. the band gap energy E$_g$), and the
heavy-holes (hh) and light-holes (lh) energies ($\mathcal{H}_{vbm}$) \cite{Leger2007,Besombes2014}. To describe the influence of a reduced symmetry
of the QD on the valence band, we considered here the four lowest energy hole states $|J,J_z\rangle$ with angular momentum $J=3/2$. A general form of
Hamiltonian describing the influence of shape or strain anisotropy on the valence band structure can be written in the basis
($|\frac{3}{2},+\frac{3}{2}\rangle,|\frac{3}{2},+\frac{1}{2}\rangle,|\frac{3}{2},-\frac{1}{2}\rangle,|\frac{3}{2},-\frac{3}{2}\rangle$) as:

\begin{equation}\label{Hvbm}
\mathcal{H}_{vbm} = \left(
\begin{array}{cccc}
0                               &s                                           &r                                   &0\\
s^*                             &\Delta_{lh}                                 &0                                   &r\\
r^*                             &0                                           &\Delta_{lh}                         &-s\\
0                               &r^*                                         &-s^*                                &0\\
\end{array}\right)
\end{equation}

Here, $r$ describes the heavy-hole / light-hole mixing induced by an anisotropy in the (xy) plane
of the QD plane and $s$ takes into account an asymmetry in the plane containing the QD growth axis
$z$. The reduction of symmetry can come from the shape of the QD (Luttinger Hamiltonian) or the
strain distribution (Bir and Pikus Hamiltonian). $\Delta_{lh}$ is the splitting between $lh$ and
$hh$ which is controlled both by the in-plane biaxial strain and the confinement.

Considering only an in-plane anisotropy ($s=0$), it follows from (\ref{Hvbm}) that the valence band mixing couples the heavy-holes $J_z=\pm3/2$ and
the light-holes $J_z=\mp1/2$ respectively. For such mixing, the isotropic part of the short range exchange interaction, which can be written in the
form $2/3\delta_0^{sr}(\overrightarrow{\sigma}.\overrightarrow{J})$, couples the two bright excitons. This mixing is also responsible for a weak
$z$-polarized dipole matrix element of the dark excitons coming from the light-hole part of the hole wave function.

A deformation in a vertical plane ($s$ term) couples the heavy-holes $J_z=\pm3/2$ and the light-holes $J_z=\pm1/2$ respectively. In this case, the
short range electron-hole exchange interaction couples $|+1\rangle$ and $|+2\rangle$ exciton on one side and $|-1\rangle$ and $|-2\rangle$ exciton on
the other side.

For a general description and as it was observed in Mn-doped QDs
\cite{Besombes2005,Trojnar2013,Besombes2014}, we can also take into account the perturbation of the
wave function of the exciton in the initial state of the optical transition by the hole-Cr exchange
interaction. This perturbation depends on the value of the exchange energy between the Cr and hole
spins and can be represented, using second order perturbation theory, by an effective spin
Hamiltonian \cite{Besombes2005,Trojnar2013,Besombes2014}

\begin{eqnarray}
{\cal H}_{scat}=-\eta S_z^2
\end{eqnarray}

\noindent with $\eta>0$.

\begin{table}[t] \centering
\caption{Values of the Cr-doped QD parameters used in PL intensity map presented in Figure\ref{FigcalcB}. The value of the parameters not listed in the table is 0.}
\begin{tabular}{p{1cm}p{1cm}p{1cm}p{1cm}p{1cm}p{1cm}p{1cm}p{1cm}}
\hline\hline
I$_{eCr}$ & I$_{hCr}$ & $\delta_0$ & $\delta_1$ & $\delta_{12}$ & $\delta_{11}$   &$\frac{|s|}{\Delta_{lh}}$  &  $\frac{|r|}{\Delta_{lh}}$             \\
$\mu eV$  & $\mu eV$  & $meV$      & $\mu eV$   & $\mu eV$      & $\mu eV$        &                           &                                        \\
\hline
-30       & 220       & -800       & 200        & 200           & 50              &0.05                       & 0.05                                   \\
\hline\hline
$arg(r)$ &  $D_0$ & $g_{Cr}$ & $g_{e}$ & $g_{h}$ & $\gamma$      &$\eta$          &$T_{eff}$                                                           \\
         & $meV$  &          &         &         & $\mu eV/T^2$  &$\mu eV$        &$K$                                                                 \\
\hline
$-\pi/2$ & 2.2    & 2        & -1      & 0.4     & 1.5           &25              &20                                                                  \\
                                                                                                              
\end{tabular}
\label{paraQD}
\end{table}

\begin{figure}[hbt]
\begin{center}
\includegraphics[width=4.0in]{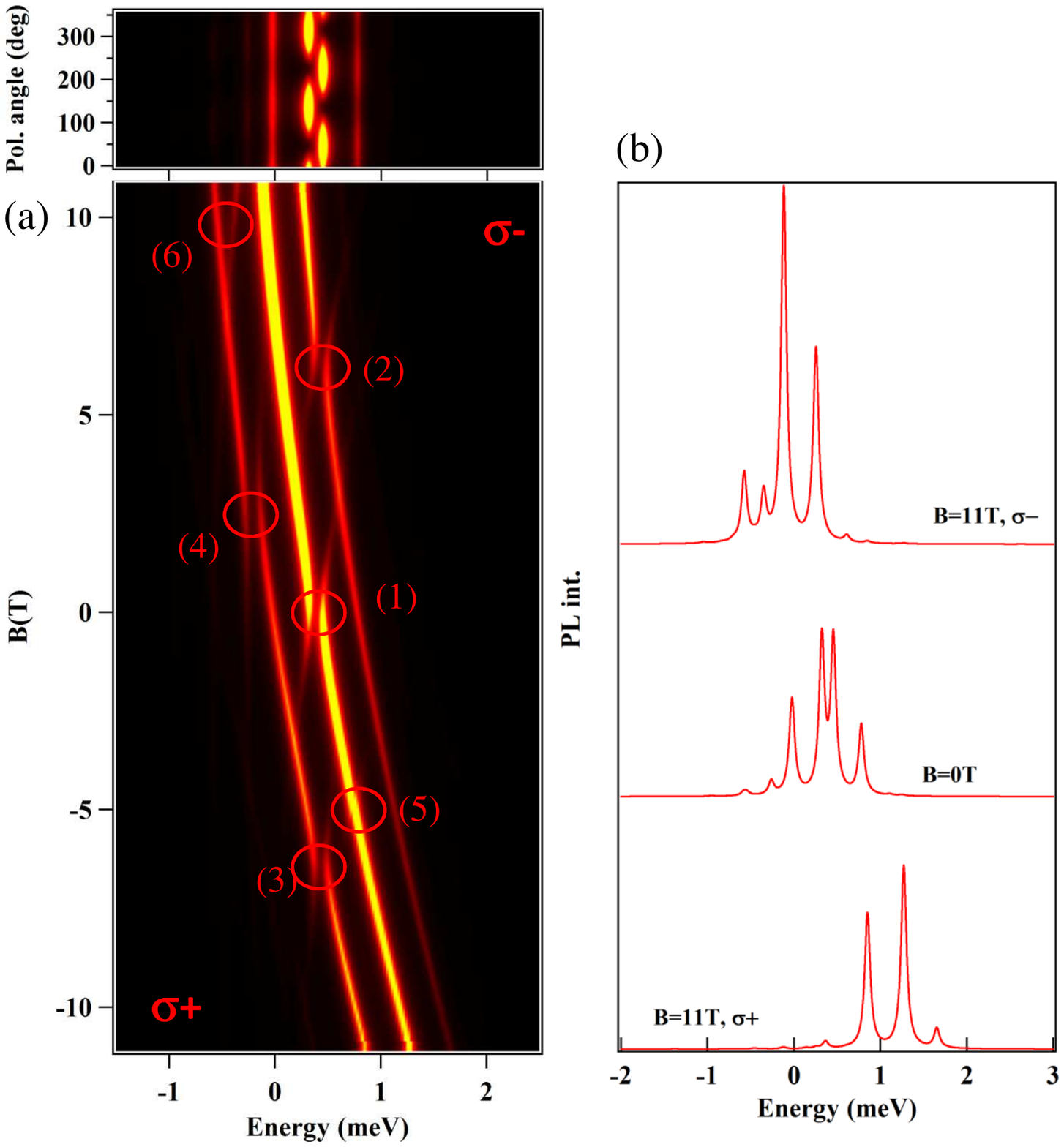}
\end{center}
\caption{(a) top panel: Calculated linear polarization PL intensity map
of X-Cr at zero field and bottom panel: calculated magnetic field dependency of the circularly polarized PL of X-Cr.
Parameters used in the model are listed in table \ref{paraQD}. (b) Calculated circularly polarized PL spectra.}
\label{FigcalcB}
\end{figure}

Using the Hamiltonian of the excited state ${\cal H}_{X-Cr}$ and the Hamiltonian of the ground state

\begin{eqnarray}
{\cal H}_{Cr}={\cal H}_{Cr,\varepsilon}+g_{Cr}\mu_B\overrightarrow{B}.\overrightarrow{S}
\end{eqnarray}

\noindent we can compute the spectrum of a QD containing a Cr atom. The occupation of the X-Cr
levels is described by an effective spin temperature T$_{eff}$ and the optical transitions
probabilities are obtained calculating the matrix elements $|\langle S_z|X,S_z\rangle|^2$ where $X$
and S$_z$ stands for the 8 possible exciton states (4 $hh$ excitons and 4 $lh$ excitons) and the Cr
spin respectively. The resulting PL spectra calculated with the parameters listed in table
\ref{paraQD} are presented in Figure\ref{FigcalcB}. The PL of X-Cr at zero field and its evolution
in magnetic field can be qualitatively reproduced. In particular, the description of the spin
states occupation by T$_{eff}$ is sufficient to reproduce the observed emission from the three low
energy X-Cr levels (Cr spin states S$_z$=0 and S$_z$=$\pm$1). The splitting of the central line at
zero field (anti-crossing (1)) and the anti-crossings under magnetic field (anti-crossings (2) and
(3) around B$_z$=6T for the Cr spin states S$_z$=+1 and anti-crossings (4) with the dark exciton
around B$_z$=2T) are also well reproduced by the model.

This model also predicts an anti-crossing around B$_z$=5T, noted (5) in Figure\ref{FigcalcB},
caused by an electron-Cr flip-flop which is not observed in the experiments. Its position is
controlled by $D_0$ and its width by $I_{eCr}$. A low value of $I_{eCr}$ is required to keep the
width of the anti-crossing smaller than the width of the PL lines. Finally an additional
anti-crossing, labeled (6), appears in the model at high magnetic field in $\sigma-$ polarization.
It is due to the mixing of bright and dark excitons associated with the same spin state of the Cr,
in this case S$_z$=0. Such anti-crossing is observed in some of the Cr doped QDs, like QD4 in
Figure\ref{FigBtherm}. Similar bright/dark mixing induced by the electron-hole exchange interaction
in low symmetry QDs can also be observed in non-magnetic QDs \cite{Leger2007}.

The model reproduces quite well the main features of the magnetic field dependence of Cr-doped QDs,
however the magnetic anisotropy $D_0$ cannot be precisely extracted from the comparison of the PL
spectra with the model. Nevertheless, for $D_0<2 meV$, a wide anti-crossing due to a VBM induced
hole-Cr flip-flop between the states $|S_z=-1\rangle|\Uparrow_h\uparrow_e\rangle$ and
$|S_z=0\rangle|\Downarrow_h\uparrow_e\rangle$ would appear for $B_z < 11T$ on the central line in
$\sigma-$ polarization. $D_0$ larger than 4 meV would push the $\pm1$ spin states at high energy
and leads to a PL intensity on the central line (S$_z$=0) much larger than observed experimentally.

\section{Resonant optical control and spin dynamics of an isolated Cr atom.}

A Cr atom presents a large spin to strain coupling particularly promising for the realization of $qubits$ in nano-mechanical systems. For a practical
use of this single spin system, one has to be able to prepare it and probe its dynamics.

\subsection{Resonant optical pumping of a single Cr spin}

\begin{figure}[hbt]
\begin{center}
\includegraphics[width=3.5in]{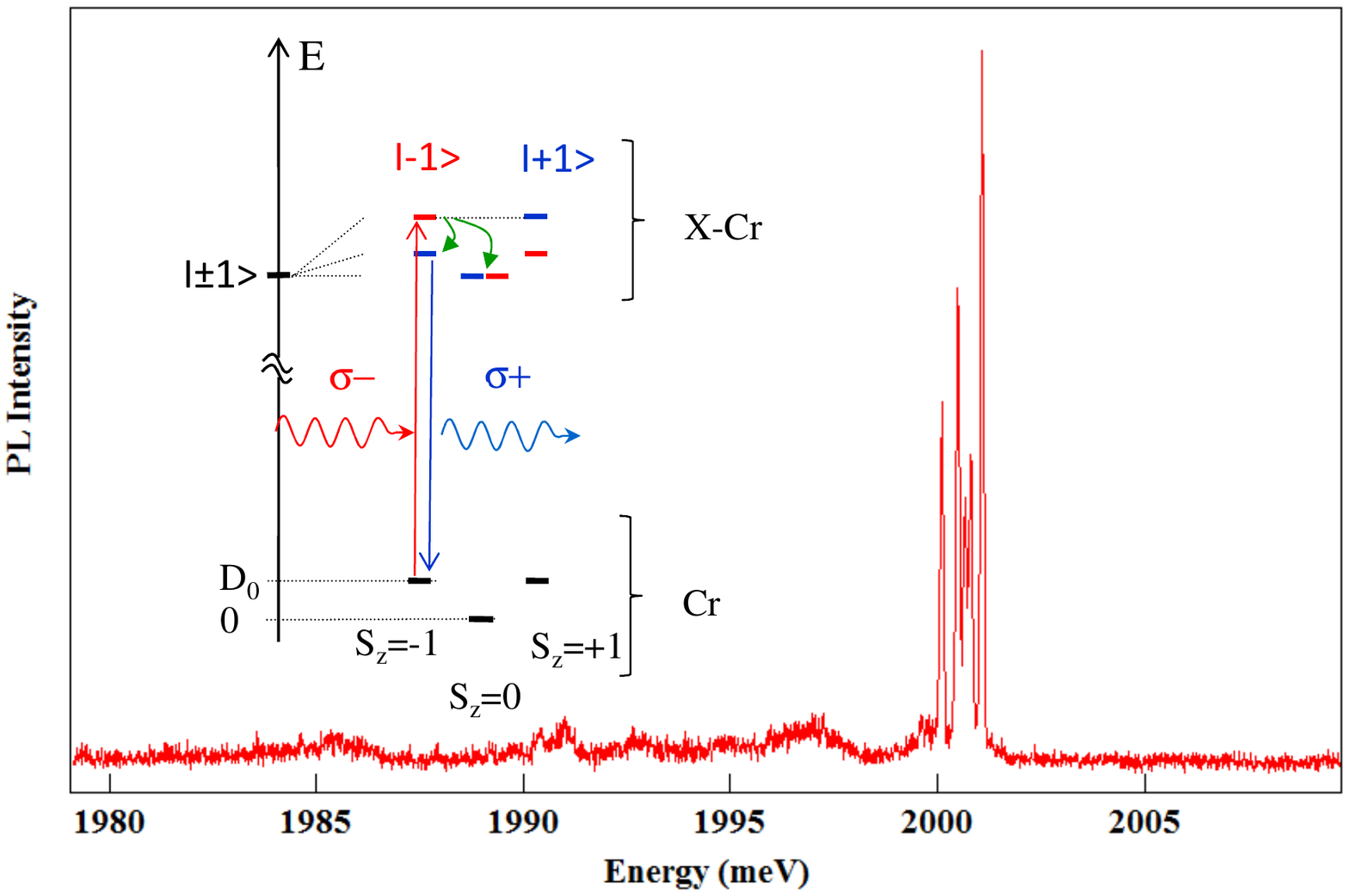}
\end{center}
\caption{Large spectral range PL spectra of QD3 showing that no contribution of the charged exciton are observed. The inset presents a scheme of the energy levels in a Cr doped QD and
the configuration of the excitation/detection in resonant pumping experiments.}
\label{Figpumplarge}
\end{figure}

To initialize and read-out the Cr spin, we developed a two wavelengths pump-probe experiment. A
circularly polarized single mode laser (\emph{resonant pump}) tuned on a X-Cr level is used to pump
the Cr spin (\emph{i.e.} empty the Cr spin state under resonant excitation). Then, a second laser,
tuned on an excited state of the QD (\emph{quasi-resonant probe}), injects excitons independently
of the Cr spin $S_z$ and drives the Cr to an effective spin temperature where the three ground
states $S_z$=0,$\pm$1 are populated \cite{Lafuente2016}. By recording the PL of a X-Cr lines in
circular polarization under this periodic sequence of excitation, we can monitor the time evolution
of the population of a given Cr spin state (see Figure\ref{Figpumplarge} for the configuration of
the excitation and detection).

\begin{figure}[hbt]
\begin{center}
\includegraphics[width=3.5in]{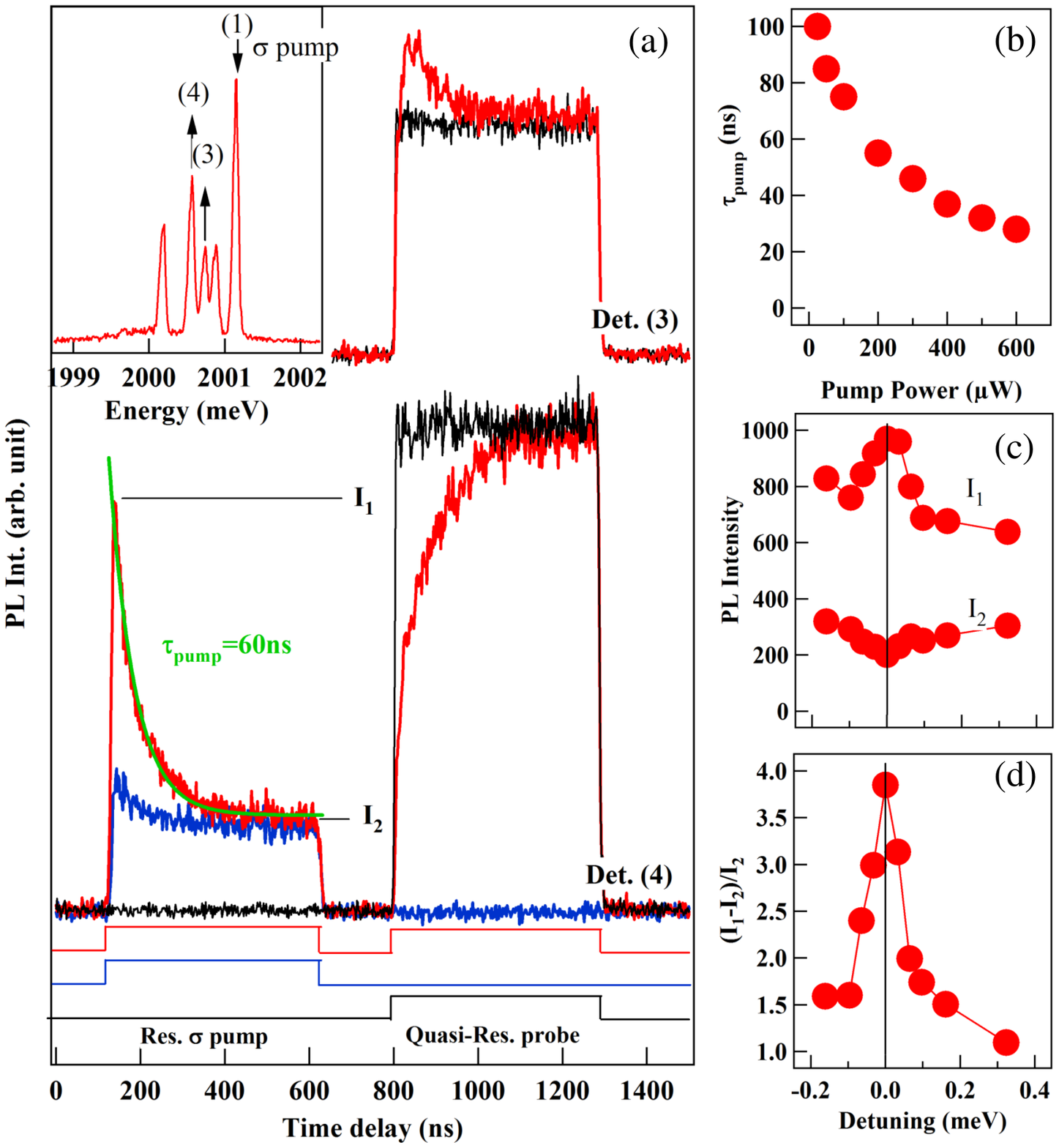}
\end{center}
\caption{(a) PL transients recorded in circular polarization on lines (4) and (3) under the resonant (pump on (1)) and quasi-resonant (probe at E$_{exc.}\approx$ 2068 meV)
optical excitation sequences displayed at the bottom. Inset: PL of X-Cr and configuration of the resonant excitation and detection. (b) Excitation power dependence of the
optical pumping time. (c) and (d): Energy detuning dependence of resonant PL intensity (I$_1$, at the beginning and I$_2$, at the end of the pump pulse) and of the corresponding
 normalized amplitude of pumping transient (I$_1$-I$_2$)/I$_2$.}
\label{Fig2WL}
\end{figure}

The main features of the optical pumping experiment performed on a neutral QD (QD3 in
Figure\ref{Figpumplarge}) are presented in Figure~\ref{Fig2WL}(a). The QD is excited on the high
energy state of X-Cr with $\sigma-$ photons (X-Cr state $|S_z=-1,-1\rangle$). This excitation can
only create an exciton in the dot if the Cr spin is $S_z$=-1. An absorption followed by possible
spin-flips of the Cr in the exchange field of the exciton progressively decreases the population of
$S_z$=-1. After this pumping sequence, the resonant pump is switched off and followed by the
non-resonant probe.

A clear signature of the optical pumping appears in the time evolution of the PL intensity of the
low energy bright exciton line (4). The PL of this line during the probe pulse, recorded in
opposite circular polarization with the resonant pump, depends on the population of $S_z$=-1. It
strongly differs between the two pump-probe sequences where the resonant pump is on or off. The
difference of intensity at the beginning of the probe pulse is a measurement of the efficiency of
the pumping. The PL intensity transient during the probe pulse corresponds to a destruction of the
non-equilibrium population distribution prepared by the pump. The speed of this spin heating
process depends on the intensity of the probe laser. As expected for an increase of the Cr spin
temperature, the population of the ground spin state $S_z$=0 also decreases during the probe pulse.
This decrease directly appears in the time evolution of the amplitude of the central X-Cr lines
during the probe pulse (Detection of line (3) in Figure~\ref{Fig2WL}(a)). The increase of the
population of $S_z$=0 during the probe pulse shows that the population of $S_z$=-1 has been
partially transferred to $S_z$=0 during the resonant pumping sequence.

A more direct way to probe the optical pumping speed and efficiency is to monitor the time
evolution of the PL during the resonant excitation by the pump pulse. Under resonant excitation on
the high energy X-Cr line, an exciton spin-flip with conservation of the Cr spin can produce a PL
on the low energy line \cite{LeGall2010}. In this process, the exciton flips its spin by emitting
(or absorbing) an acoustic phonon. Such spin-flip is enhanced by the large acoustic phonon density
of states at the energy of the inter-level splitting induced by the exchange interaction with the
Cr spin which act as an effective magnetic field \cite{Tsitsishvili2003,Roszak2007}. The resulting
weak resonant PL signal depends on the occupation of the Cr state $S_z$=-1 and is used to monitor
the time dependence of the spin selective absorption of the QD.

The time evolution of the PL of the low energy line of X-Cr (line (4)) under an excitation on the
high energy line (1) is presented in Figure~\ref{Fig2WL}(a) for two different pump-probe sequences:
probe on and probe off. When the probe laser is on, a large effective Cr spin temperature is
established before each pumping pulse. The amplitude of the resonant PL is maximum at the beginning
of the pump pulse ($I_1$) and progressively decreases. A decrease of the amplitude of about 80\% is
observed after a characteristic time in the tens of $ns$ range. As expected for a spin optical
pumping process, the characteristic time of the PL transient decreases with the increase of the
pump laser intensity (Figure~\ref{Fig2WL}(b)) \cite{LeGall2010}. When the probe laser is off, the
initial amplitude of the PL transient during the pump pulse is significantly weaker. This decrease
is a consequence of the conservation of the out of equilibrium Cr spin distribution during the dark
time between two consecutive pumping pulses.

The steady state resonant PL intensity reached at the end of the pump pulse ($I_2$) depends on the
optical pumping efficiency which is controlled by the ratio of the spin-flip rate for the Cr spin
in the exchange field of the exciton and the relaxation of the Cr spin in the absence of carriers
in the dot. However, even with cross-circularly polarized excitation/detection, this steady state
PL can also contain a weak contribution from an absorption in the acoustic phonon sideband of the
low energy line (4) \cite{Besombes2001}. Figure~\ref{Fig2WL}(c) presents the amplitude of the
resonant PL detected on line (4) for a detuning of the pump around the high energy line (1). A
resonance is observed in the initial amplitude $I_1$ of the PL. It reflects the energy and
excitation power dependence of the absorption of the QD. A small decrease of the steady state PL
$I_2$ is also observed at the resonance. As displayed in Figure~\ref{Fig2WL}(d), the corresponding
normalized amplitude of the pumping transient, $(I_1-I_2)/I_2$, presents a clear resonant behavior
demonstrating the excitation energy dependence of the optical pumping process. The width of the
resonance ($\sim 100\mu eV$) is the width of the QD's absorption broadened by the fluctuating
environment \cite{Sallen2011}.

\subsection{Cr spin relaxation in the dark}

\begin{figure}[hbt]
\begin{center}
\includegraphics[width=3.5in]{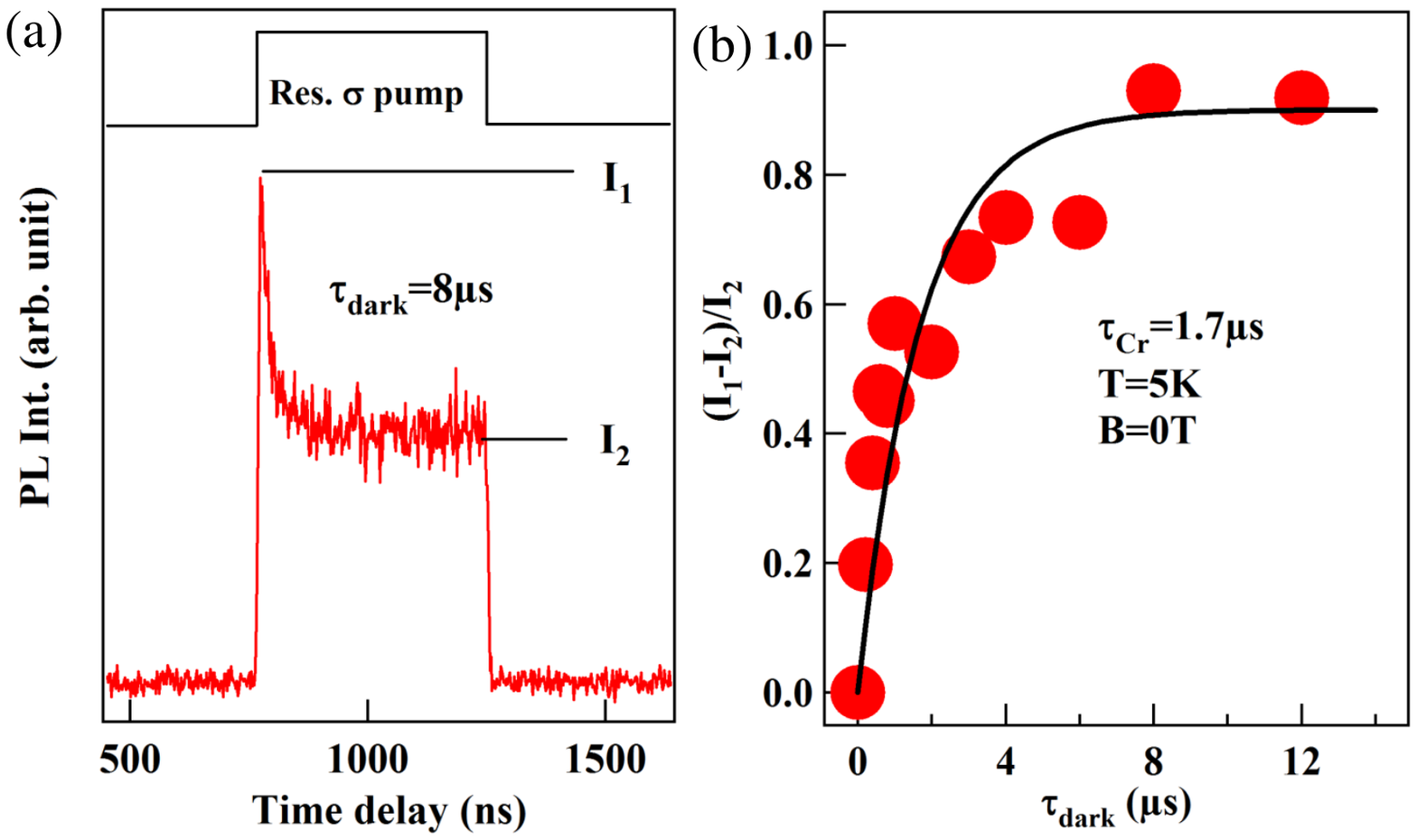}
\end{center}
\caption{(a) Time evolution of the PL intensity of line (4) of X-Cr under resonant excitation on line (1) with a circularly polarized excitation pulse.
(b) Evolution of the amplitude of the pumping transient $(I_1-I_2)/I_2$ as a function of the dark time $\tau_{dark}$ between the excitation pulses.
The black line is an exponential evolution with a characteristic time $\tau_{Cr}=1.7\mu s$}
\label{FigRelax}
\end{figure}

With this resonant optical pumping technique used to prepare and read-out the Cr spin, we performed
pump-probe experiments to observe its relaxation time in the absence of carriers
(Figure~\ref{FigRelax}). A non-equilibrium distribution of the Cr spin population is prepared with
a circularly polarized resonant pump pulse on the high energy X-Cr line (1). The pump laser is then
switched off, and switched on again after a dark time $\tau_{dark}$. The amplitude of the pumping
transient observed on the resonant PL of the low energy line (4) depends on the Cr spin relaxation
during $\tau_{dark}$. As presented in Figure~\ref{FigRelax}(b), the amplitude of the transient is
fully restored after a dark time of about 10 $\mu$s showing that after this delay the Cr spin is in
equilibrium with the lattice temperature (T=5K). Let us note, however, that the initial amplitude
of the pumping transient in this case is weaker than the one observed after a non-resonant probe
pulse (Figure~\ref{FigRelax}(a)). This means that the non-resonant optical excitation drives the Cr
spin to an effective temperature much larger than the lattice temperature.

These measurements reveal a significantly different Cr spin-flip times under optical excitation (tens of nanosecond range) and in the dark
(microseconds range). The fast Cr spin-flip under optical excitation can be due to the interaction with carriers (exchange induced Cr spin flips
\cite{Cao2011}) but can also be induced by the interaction with non-equilibrium acoustic phonons created during the energy relaxation of the injected
carriers. Both mechanisms probably contribute to the Cr spin heating.

\subsection{Dynamics under optical excitation and mechanism of optical pumping for a Cr spin}

In order to identify the main mechanism responsible for the optical pumping, we analyzed the PL
intensity distribution under resonant excitation \cite{Lafuente2018}. We have seen that under
circularly polarized resonant excitation of the high energy X-Cr line (1), a weak PL is observed in
cross-circular polarization on the low energy bright exciton line (4) after a spin-flip of the
exciton conserving the Cr spin. In this experimental configuration the same spin state of the Cr is
excited and detected (see the energy level diagram presented in Figure \ref{Figpumplarge}(a)). The
intensity transient observed during the resonant PL directly reflects the pumping of the Cr spin
state under excitation (either $S_z=+1$ or $S_z=-1$).

\begin{figure}[hbt]
\begin{center}
\includegraphics[width=4.0in]{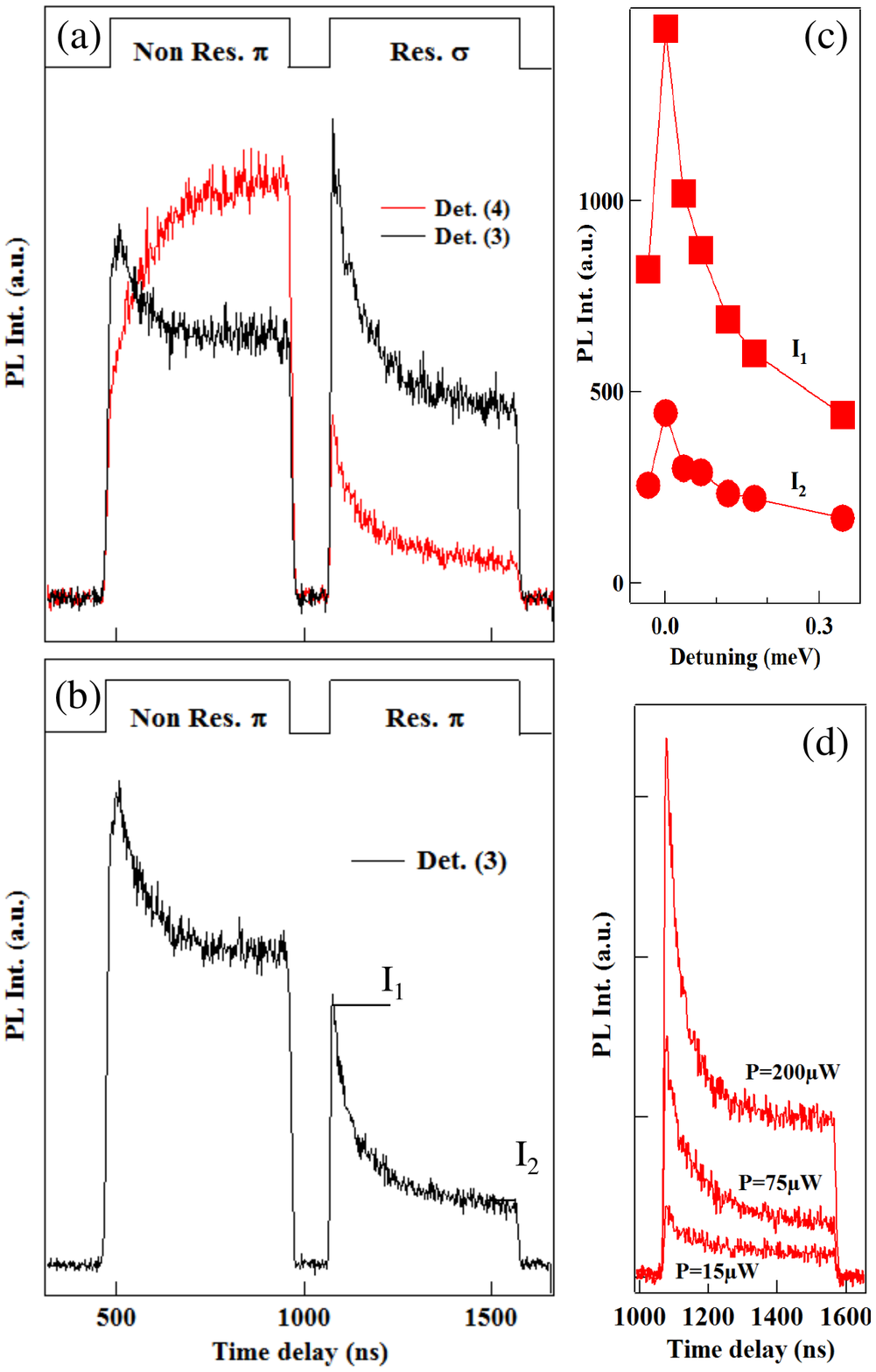}
\end{center}
\caption{(a) and (b): resonant optical pumping experiments for cross circularly polarized excitation/detection (a) and cross-linearly polarized excitation/detection (b).
(c) Dependence on the energy detuning around line (1) of the resonant PL intensity detected on the central line (3) for cross-linearly polarized excitation/detection. (d) Excitation power dependence of the resonant PL
transient detected on line (3) for cross-linearly polarized excitation/detection.}
\label{FigFluoRes}
\end{figure}

However, as presented in Figure\ref{FigFluoRes}(a), during this resonant pumping process a
luminescence is also observed on the central lines (2) and (3) (linearly polarized bright excitons
resulting from the mixing of $|0\rangle|\Uparrow_h\downarrow_e\rangle$ and
$|0\rangle|\Downarrow_h\uparrow_e\rangle$). This resonant PL presents a transient with a similar
time scale and amplitude as the one detected on the low energy line (4). The steady state level of
the resonant PL detected on $S_z=0$ is however much lager than for a detection on the low energy
bright exciton. This steady state PL intensity is strongly reduced under cross linearly polarized
excitation and detection (figure \ref{FigFluoRes}(b)).

The central line (2) and (3) are linearly polarized along two orthogonal directions. For a
circularly polarized excitation on the high energy side of these transitions and for a
cross-circularly polarized detection, a weak absorption in the acoustic phonon sideband
\cite{Besombes2001} is expected to induce some resonant PL. However, the presence of the intensity
transient due to the pumping of the resonantly excited spin states $S_z=\pm1$ shows that there is
also an efficient population transfer between the high energy bright excitons
($|+1\rangle|\Uparrow_h\downarrow_e\rangle$ or $|-1\rangle|\Downarrow_h\uparrow_e\rangle$) created
during the resonant optical excitation and the lower energy bright exciton states
($|0\rangle|\Downarrow_h\uparrow_e\rangle$ or $|0\rangle|\Uparrow_h\downarrow_e\rangle$).

For a cross-linearly polarized excitation and detection, the absorption in the acoustic phonon
side-band is strongly reduced. The pumping transient resulting from an absorption on the states
$S_z=\pm1$ and a transfer toward $S_z=0$ dominates the resonant PL. This spin-flip transfer is
confirmed by the excitation energy and power dependence of the resonant PL signal. The PL intensity
on line (3) presents a resonant behavior when scanning the pump laser around the high energy line
(1) (Figure \ref{FigFluoRes}(c)). As expected for an optical pumping process, the time scale of the
PL transient is also significantly reduced with the increase of the excitation intensity (Figure
\ref{FigFluoRes}(d))).

This efficient transfer towards $S_z=0$ during the resonant optical pumping process is also
observed in the evolution of the PL intensity during the non-resonant probe (heating) pulse. As
expected after a resonant pumping of the spin states $S_z=\pm1$, their population re-increase
during the non-resonant heating pulse. Simultaneously, a significant decrease of the intensity of
the line (3) is observed both for a circularly or a linearly polarized detection (figure
\ref{FigFluoRes}(a) and (b)). This decrease corresponds to a decrease of the population of $S_z$=0
during the heating process showing that a significant part of the Cr spin population has been
transferred to $S_z$=0 during the resonant optical pumping.

\begin{figure}[hbt]
\begin{center}
\includegraphics[width=3.0in]{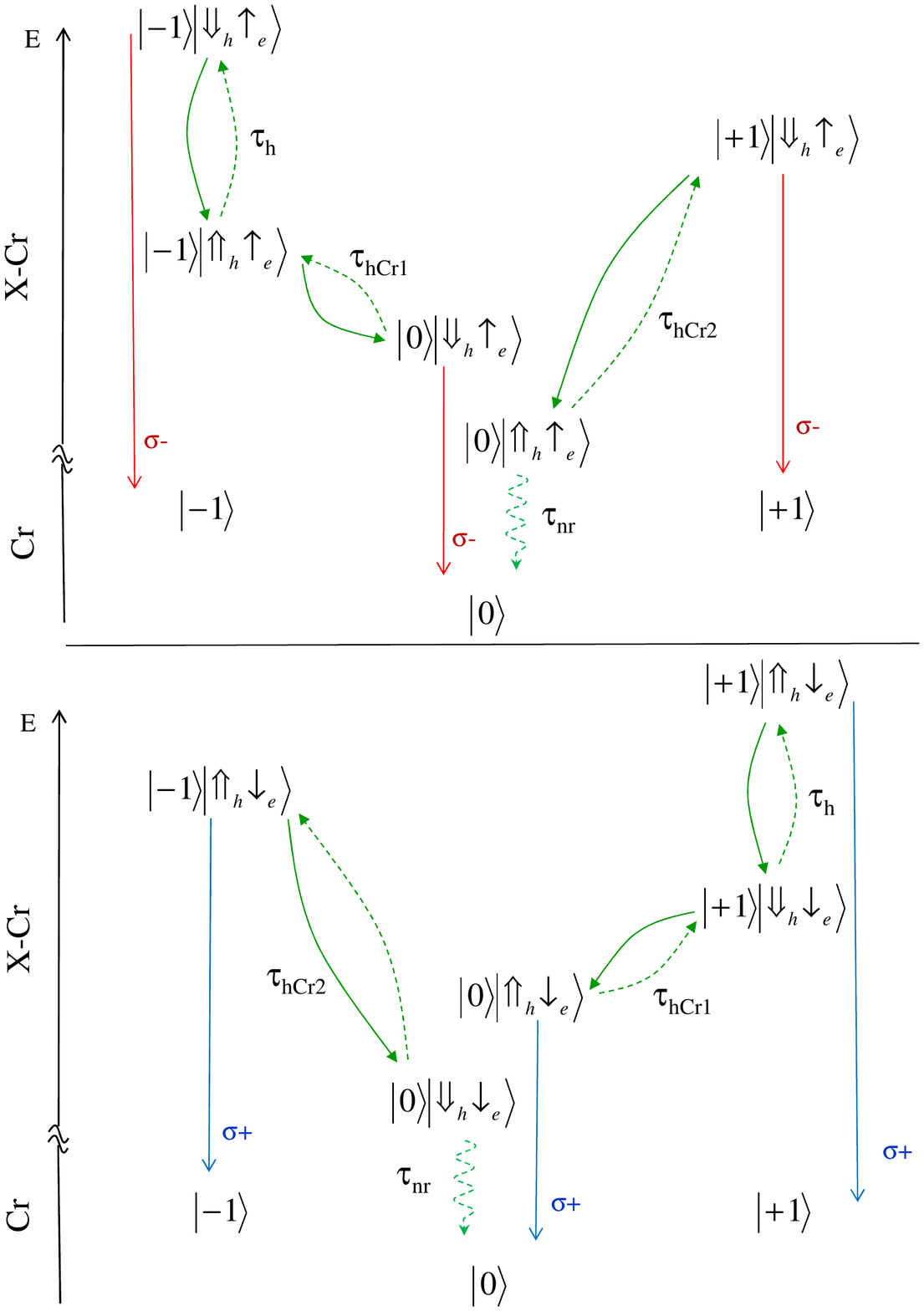}
\end{center}
\caption{Spin relaxation paths within the exciton-Cr complex (X-Cr) for an excitation of the bright exciton $|\Downarrow_h\uparrow_e\rangle$ in $\sigma-$
polarization (top) and of the bright exciton $|\Uparrow_h\downarrow_e\rangle$ in $\sigma+$ polarization (bottom). $\tau_h$ is the spin flip time of the hole,
 $\tau_{hCri}$ is a hole-Cr flip-flop times and $\tau_{nr}$ is the non-radiative recombination time of the dark excitons.}
\label{Fighspin}
\end{figure}

A fast hole-Cr flip-flop can explain an efficient transfer of population from the exciton-Cr levels
S$_z=\pm1$ towards the low energy states S$_z=0$. The proposed relaxation path is illustrated in
the energy level scheme presented in figure \ref{Fighspin} which displays the possible relaxation
channels involving a hole-Cr flip-flop for an initial excitation of the bright excitons coupled
with $S_z=\pm1$.

Starting from the low energy bright excitons $|+1\rangle|\Downarrow_h\uparrow_e\rangle$ or $|-1\rangle|\Uparrow_h\downarrow_e\rangle$, a hole-Cr
flip-flop with conservation of the electron spin can directly transfer the population towards the dark excitons associated with S$_z=0$ with a spin
flip time $\tau_{hCr2}$. With the same mechanism, a direct transfer is possible from the low energy dark excitons
$|+1\rangle|\Downarrow_h\downarrow_e\rangle$ or $|-1\rangle|\Uparrow_h\uparrow_e\rangle$, to the bright excitons associated with S$_z=0$ with a spin
flip time $\tau_{hCr1}$ (See figure \ref{Fighspin}).

Starting from the high energy bright excitons $|+1\rangle|\Uparrow_h\downarrow_e\rangle$ or $|-1\rangle|\Downarrow_h\uparrow_e\rangle$, a hole-Cr
flip-flop is unlikely as it would involve a transfer toward the high energy Cr spin states $S_z=\pm2$. However, a spin flip of the hole with
characteristic time $\tau_h$ conserving the spin of the Cr can induce a transfer to the low energy dark excitons
$|+1\rangle|\Downarrow_h\downarrow_e\rangle$ or $|-1\rangle|\Uparrow_h\uparrow_e\rangle$. Such spin-flip from a bright to a dark exciton can occur in
QDs in a time-scale of a few nanoseconds \cite{Cao2011}. A hole-Cr flip-flop in a timescale $\tau_{hCr1}$ can then transfer the population from these
dark excitons to the S$_z$=0 bright exciton which then recombines optically (See figure \ref{Fighspin}).

From the hight energy dark excitons $|+1\rangle|\Uparrow_h\uparrow_e\rangle$ or $|-1\rangle|\Downarrow_h\downarrow_e\rangle$ a hole-Cr flip-flop is
also unlikely (transfer toward the high energy Cr spin states $S_z=\pm2$). However these states are coupled by a hole spin-flip to the nearby low
energy bright excitons $|+1\rangle|\Downarrow_h\uparrow_e\rangle$ or $|-1\rangle|\Uparrow_h\downarrow_e\rangle$ and can then follow this path to be
transferred towards $S_z=0$.

The spin relaxation channels involving a transfer from the dark exciton with $S_z=\pm1$ to the
lower energy bright excitons $S_z=0$ are made irreversible by the fast ($\approx$ 250 ps) radiative
recombination of the final low energy exciton state. This unusual situation with some of the bright
excitons at lower energy than the dark ones can induce an out of equilibrium distribution on the Cr
spin states with an enhanced population of the S$_z$=0 ground state. The transfer mechanism
involving a hole-Cr flip-flop is enhanced by the increase of the probability of presence of an
exciton in the QD. This can be observed in the excitation power dependence of the distribution of
the non-resonant PL intensity \cite{Lafuente2018}.

Under resonant excitation on the hight energy levels of X-Cr (line (1)), a spin flip of the hole
followed by a fast hole-Cr flip-flops can explain the observed transfer of excitation toward the
bright excitons associated with $S_z=0$. This process which efficiently changes the Cr spin is the
likely to be the main source for the resonant optical pumping demonstrated in Cr-doped QDs
\cite{Lafuente2017Cr}.

\subsection{Phonon induced hole-Cr flip-flops}

The time-scale of the hole-Cr flip-flops in a Cr-doped QD induced by the interaction with the
continuum of bulk acoustic phonons can be estimated using the Fermi golden rule. The spin-flip
process that we consider here is based on the interplay of the hole-magnetic atom exchange
interaction and the interaction with the strain field of acoustic phonons \cite{Lafuente2017Mn}.
Similar models, combining exchange interaction and coupling with acoustic phonons, where developed
to explain the exciton spin relaxation in QDs \cite{Tsitsishvili2003,Roszak2007}.

Let us consider the two X-Cr states $|+1\rangle|\Downarrow_h;\downarrow_e\rangle$ and
$|0\rangle|\Uparrow_h;\downarrow_e\rangle$ that can be coupled via a hole-Cr flip-flop. The
non-diagonal terms of the hole-Cr exchange interaction couples the heavy-holes and light-holes
excitons levels separated in energy by $\Delta_{lh}$ through a hole-Cr flip-flop. To the first
order in $I_{hCr}/\Delta_{lh}$, the two perturbed heavy-hole exciton ground states can be written:

\begin{eqnarray}
\widetilde{|+1\rangle|\Downarrow_h;\downarrow_e\rangle}=|+1\rangle|\Downarrow_h;\downarrow_e\rangle-\frac{\sqrt{18}}{2}\frac{I_{hCr}}{\Delta_{lh}}|0\rangle|\downarrow_h;\downarrow_e\rangle\nonumber\\
\widetilde{|0\rangle|\Uparrow_h;\downarrow_e\rangle}=|0\rangle|\Uparrow_h;\downarrow_e\rangle-\frac{\sqrt{18}}{2}\frac{I_{hCr}}{\Delta_{lh}}|+1\rangle|\uparrow_h;\downarrow_e\rangle
\end{eqnarray}

\noindent where we neglect the exchange energy shifts of the exciton-Cr levels much smaller than $\Delta_{lh}$.

The strain field produced by acoustic phonon vibrations couples the perturbed exciton-Cr states through the Hamiltonian term:

\begin{eqnarray}
\label{int}
\widetilde{\langle\Downarrow_h;\downarrow_e|\langle+1|}H_{BP}\widetilde{|0\rangle|\Uparrow_h;\downarrow_e\rangle}=2\times(-\frac{\sqrt{18}}{2}\frac{I_{hCr}}{\Delta_{lh}})\times r^*
\end{eqnarray}

\noindent with $r=\sqrt{3}/2b(\epsilon_{xx}-\epsilon_{yy})-id\epsilon_{xy}$ a strain-dependent non-diagonal term of the Bir-Pikus Hamiltonian
$H_{BP}$ \cite{kp}. The coupling of these exciton-Cr states is then a result of an interplay between the hole-Cr exchange interaction and the strain
field of acoustic phonons.

In analogy with the model developed in reference (36) to describe hole-Mn flip-flops in positively charged Mn-doped QDs, the decay rate associated
with the emission of phonons which is deduced from the Fermi golden rule and the matrix element (\ref{int}) can be written:

\begin{eqnarray}
\tau^{-1}&=&\sum_{\lambda}\frac{18}{(2\pi)^2}\left(\frac{I_{hMn}}{\Delta_{lh}}\right)^2\left(\frac{\omega_0}{c_{\lambda}}\right)^3\frac{1}{2\hbar\rho c_{\lambda}^2}\frac{\pi}{4}\left(3b^2+d^2\right)\nonumber\\
&\times&\left(n_B(\omega_0)+1)\right)\int_0^{\pi}d\theta\sin\theta|\mathcal{F}_{\lambda}(\omega_0,\theta)|^2G_{\lambda}(\theta)
\label{fermi}
\end{eqnarray}

\noindent where the summation is taken over the acoustic phonon branches $\lambda$ (one longitudinal $l$ and two transverse $t_1,t_2)$) of
corresponding sound velocity c$_{\lambda}$. The geometrical form factors for each phonon branch appearing in (\ref{fermi}), $G_{\lambda}(\theta)$,
are given by $G_{l}(\theta)=\sin^4\theta$, $G_{t_1}(\theta)=\sin^2\theta\cos^2\theta$ and $G_{t_2}(\theta)=\sin^2\theta$.

\begin{figure}[hbt]
\begin{center}
\includegraphics[width=4.5in]{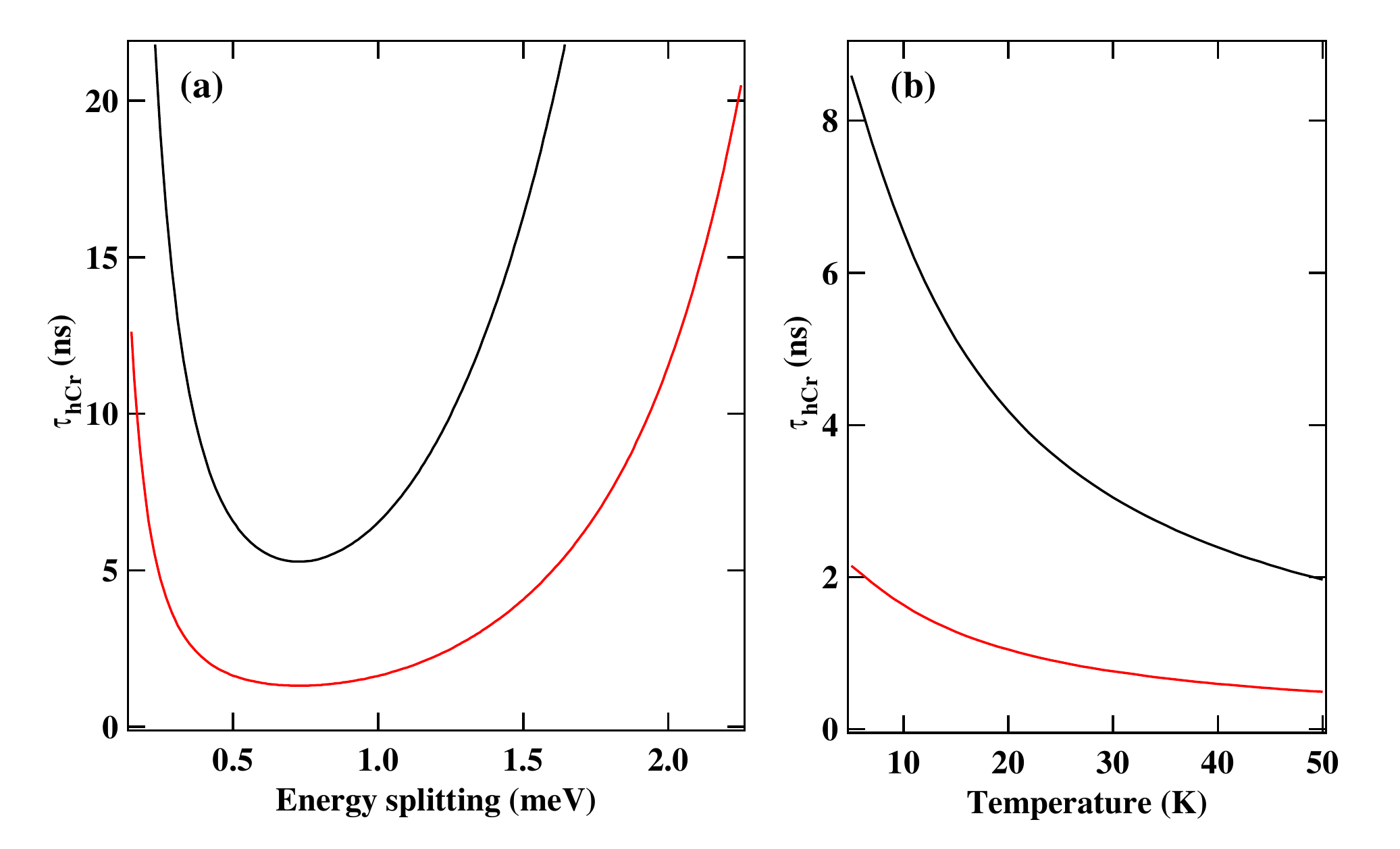}
\end{center}
\caption{(a) Relaxation time $\tau_{hCr}$ between the states $|+1\rangle|\Downarrow_h;\downarrow_e\rangle$ and $|0\rangle|\Uparrow_h;\downarrow_e\rangle$
(or $|-1\rangle|\Uparrow_h;\uparrow_e\rangle$ and $|0\rangle|\Downarrow_h;\uparrow_e\rangle$) as a function of the energy splitting between those states,
calculated at a temperature T = 10 K with a Gaussian hole wave function (parameters $l_z$ = 1,25 nm and $l_\perp$ = 3 nm) and $\Delta_{lh}$=25 meV (red),
$\Delta_{lh}$=50 meV (black). The other CdTe and QD parameters used in the calculation can be found in table \ref{paraph}.
(b) Temperature variation of the relaxation time for an energy splitting E=1 meV and $\Delta_{lh}$=25 meV (red), $\Delta_{lh}$=50 meV (black).}
\label{Figcalc}
\end{figure}

Numerical calculation of the hole-Cr spin-flip time in a CdTe QD are presented in figure
\ref{Figcalc}. We used in the calculation a Gaussian wave function for the hole with in-plane and
z-direction parameters $l_\perp$ and $l_z$ respectively, the material parameters of CdTe and
typical parameters of self-assembled CdTe/ZnTe QDs listed in table \ref{paraph}.

The calculated spin relaxation time strongly depends on the energy splitting between the X-Cr
states involved in the hole-Cr flip-flops (figure \ref{Figcalc}(a)). This dependence on the high
energy side is controlled by the size of the hole wave function which limits the wave vector of the
acoustic phonons that can interact with the hole. The calculated spin-flip time can be in the
nanosecond range for an energy splitting between 0.5 meV and 2 meV. In the studied Cr-doped QDs,
the splitting between the low energy dark excitons states
$|+1\rangle|\Downarrow_h\downarrow_e\rangle$ or $|-1\rangle|\Uparrow_h\uparrow_e\rangle$ and the
bright excitons with $S_z=0$, which are coupled by a hole-Cr flip-flop, is typically in the 1 meV
range. A spin-flip time of a few ns can then be expected.

\begin{table}[hbt] \centering
\caption{Material (CdTe) \cite{Adachi2005} and QD parameters used in the calculation of the coupled hole and Cr spin relaxation time presented in figure \ref{Figcalc}.}
\begin{tabular}{lcr}
\hline\hline
CdTe& &\\
\hline
Deformation potential constants & $|b|$       &  1.0 eV          \\
                                & $|d|$       &  4.4 eV          \\
Longitudinal sound speed        & c$_l$       &  3300 m/s        \\
Transverse sound speed          & c$_t$       &  1800 m/s        \\
Density                         & $\rho$      &  5860 kg/m$^3$   \\
\hline
Quantum dot& &\\
\hline
Cr-hole exchange energy         & I$_{hCr}$     &  0.22 meV      \\
hh-lh exciton splitting         & $\Delta_{lh}$ &  25 or 50 meV  \\
Hole wave function widths:      &               &                \\
- in plane                      & l$_{\bot}$    & 3.0 nm         \\
- z direction                   & l$_z$         & 1.25 nm        \\
\hline\hline
\end{tabular}
\label{paraph}
\end{table}

For the low energy bright excitons $|+1\rangle|\Downarrow_h\uparrow_e\rangle$ or
$|-1\rangle|\Uparrow_h\downarrow_e\rangle$ and the dark excitons states with $S_z=0$ also coupled
by hole-Cr flip-flops, their energy splitting is larger than 3 meV and the spin-flip rates are
expected to be significantly reduced. For the high energy Cr spins states, S$_z=\pm2$, the energy
splitting is typically larger than 10 meV and the corresponding flip-flops induced by the discussed
mechanism can be neglected.

The estimated hole-Cr flip-flop time is also strongly sensitive to the effective energy splitting between heavy-hole and light-hole $\Delta_{lh}$.
This simple parameter is used in our model for an effective description of the valence band mixing. It can describe complex effects such as a
coupling of the confined heavy-hole with ground state light-holes in the barriers \cite{Michler2003} or effective reduction of heavy-hole/light-hole
splitting due to a presence of a dense manifold of heavy-hole like QD states lying between the confined heavy-hole and light-hole levels
\cite{Bester2015}. This parameter can significantly change from dot to dot and modify the hole-Cr flip-flop time.

The calculated hole-Cr flip-flop rate also depends on the temperature through the stimulated
emission of acoustic phonons. For an energy splitting of 1 meV the spin flip rate involving the
emission of phonons is increased by a factor of about four by increasing the temperature from 5K to
50K (figure \ref{Figcalc}(b)). Such enhancement of the hole-Cr spin-flip rate can not only be
induced by an increase of the lattice temperature but also, as in our experiments, by the presence
of non-equilibrium phonons that are generated by the optical excitation inside or in the vicinity
of the QD.

This calculated short flip-flop time in the $ns$ range, is consistent with the initialization time measured in resonant optical pumping experiments
and controls the spin dynamics under resonant excitation.

\subsection{Optical Stark effect on an individual Cr spin}

\begin{figure}[hbt]
\begin{center}
\includegraphics[width=4.0in]{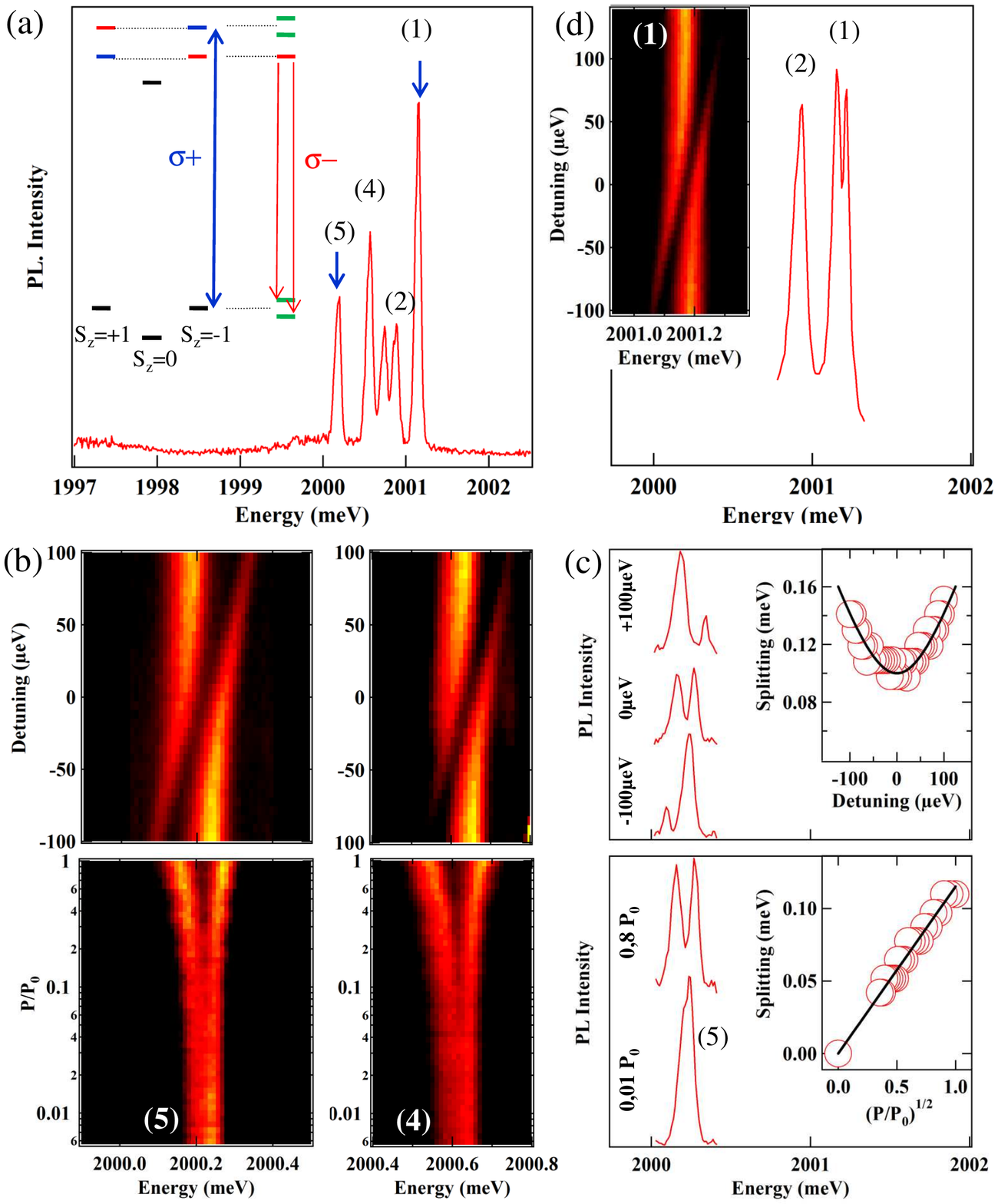}
\end{center}
\caption{(a) PL of X-Cr and configuration of excitation in the resonant optical control experiments. The inset illustrate the laser induced splittings
in the ground and excited states for a $\sigma+$ excitation on S$_z$=-1. (b) PL intensity maps of lines (5) and (4) for an excitation on (1) as a function
of the detuning (top) and of the excitation intensity (bottom). The PL is produced by a second non-resonant laser. The corresponding emission line-shapes are presented
in (c) for line (5). The insets in (c) show the splitting of the PL doublet as a function of the excitation intensity (bottom) and laser detuning (top).
The fit is obtained with $\hbar\Omega_r$= 100 $\mu eV$. (d) PL of line (1) and (2) for a laser on resonance with the dark exciton state (5). Inset: PL intensity
map of line (1) as a function of the laser detuning around (5).}
\label{FigStark}
\end{figure}

The resonant optical excitation on a X-Cr line can also be used to tune the energy of any Cr spin
state through the optical Stark effect \cite{LeGall2011,Xu2007,Muller2008}. Such energy shift could
be exploited to control the coherent dynamics of the magnetic atom \cite{Jamet2013,Reiter2013}.
This optical control technique is presented in Figure~\ref{FigStark}. When a high intensity single
mode laser is tuned to the high energy line of X-Cr in $\sigma+$ polarization (X-Cr state
$|S_z=-1,+1\rangle$), a splitting is observed in $\sigma-$ polarization in the PL of the two low
energy lines produced by a second non-resonant laser.

At high excitation intensity, a strong coupling with the resonant laser field mixes the states with
a Cr spin component $S_z$=-1 in the presence (X-Cr) or absence (Cr alone) of the exciton. In the
ground state of the QD (Cr alone) two hybrid matter-field states are created (inset of
Figure~\ref{FigStark}(a)). Their splitting,
$\hbar\Omega_r^{\prime}=\hbar\sqrt{\Omega_r^2+\delta^2}$, depends on the energy detuning of the
laser $\hbar\delta$ and on its intensity through the Rabi energy $\hbar\Omega_r$ \cite{Boyle2009}.
It can be observed in the PL of all the X-Cr states associated with $S_z$=-1: the low energy bright
exciton state $|S_z=-1,-1\rangle$ (line (4)) and the dark exciton $|S_z=-1,+2\rangle$ (line (5)),
close in energy to the bright exciton and which acquires some oscillator strength through the
exciton mixing induced by the electron-hole exchange interaction in a low symmetry QD
\cite{Lafuente2016}.

The splitting measured on line (5) for a resonant excitation on line (1) is plotted as a function
of the square root of the resonant laser intensity in Figure~\ref{FigStark}(c) and shows that, as
expected for a two level system, it linearly depends on the laser field strength. The Rabi
splitting can reach 150 $\mu eV$ at high excitation power. As the pump laser is detuned, the
optically active transitions asymptotically approaches the original excitonic transitions where the
remaining offset is the optical Stark shift.

A resonant laser permits to address any spin state of the Cr and selectively shift its energy. For
instance, as presented in Figure~\ref{FigStark}(d), a $\sigma$+ excitation on the dark exciton
state (5) induces a splitting of the high energy line (1) in $\sigma$- polarization (state
$|S_z=+1,-1\rangle$) without affecting the central line (2). This shows that such resonant
excitation can be used to tune the energy of $S_z$=+1 without affecting $S_z$=0. The energy tuning
induced by a coherent optical driving is particularly interesting for the control of the Cr spin
states $S_z$=$\pm$1. These states could be efficiently mixed by applied weak anisotropic in-plane
strain through a fine structure term of the form $E(S_x^2-S_y^2)$ \cite{Lafuente2016}. A relative
shift of the energy of $S_z$=+1 or $S_z$=-1 by a resonant optical excitation would affect their
coupling and consequently the Cr spin coherent dynamics.

Future applications of Cr as a spin $qubit$ in hybrid nano-mechanical systems
\cite{Pigeau2015,Barfuss2015} will exploit the efficient mixing of the Cr spin states $S_z$=+1 and
$S_z$=-1 induced by anisotropic in-plane strain. The resulting mixed spin states, together with an
exciton, form an optical three level $\Lambda$ system. A coherent optical driving of this level
structure opens the possibility of using coherent spectroscopy techniques such as coherent
population trapping \cite{Houel2014} for a sensitive probing of the splitting of the $S_z$=+1 and
$S_z$=-1 induced by the local strain at the atom location. The strain field that could be probed
with this technique will depend on the coherence time of the Cr $\{+1;-1\}$ spin $qubit$.

\section{Perspectives: a Cr atom as a spin qubit in hybrid spin-mechanical systems.}

It has been shown recently that an individual spin can be used as a $qubit$ to interact with the
motion of a nano-mechanical oscillator and probe its position, cool it down or prepare
non-classical state of the mechanical motion \cite{Lee2017}. It has been also demonstrated
theoretically that the coupling of two or more localized spins with the same mode of a mechanical
oscillator could be used to induce long range coherent coupling between spins \cite{Bennett2013}, a
usually difficult task to achieve with spins in solid state systems. Among mechanical systems,
Surface Acoustic Wave (SAW) {\it i.e.} phonons propagating at the surface of a solid, are
particularly promising. SAW can coherently propagate over long distances at the surface of a solid
state device, they can be guided or confined in acoustic wave-guide or acoustic cavities and they
can interact with different kinds of $qubits$. They are in particular proposed as efficient quantum
bus between dissimilar $qubits$ \cite{Schuetz2015,Lemonde2018}. In the case of spin $qubits$, the
performance of SAW based nano-mechanical devices will be enhanced by the use of localized spins
with large intrinsic spin to strain couplings. Cr is in that sense particularly promising. As we
have seen in section 2.1, in-plane strain oriented along the [110] direction ($\epsilon_{xy}$) will
be strongly coupled to the spin degree of freedom of a Cr.

\begin{figure}[hbt]
\begin{center}
\includegraphics[width=4.5in]{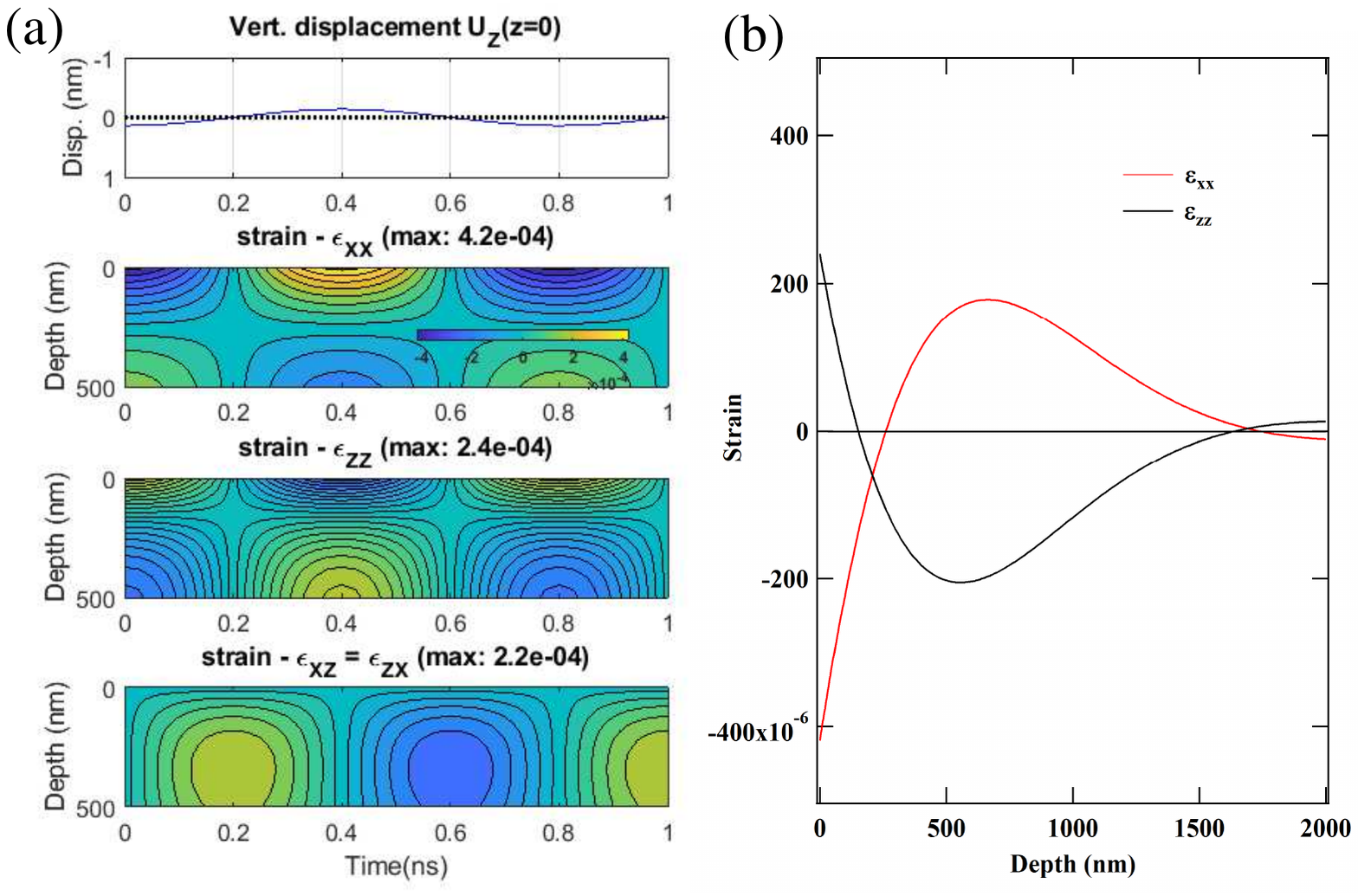}
\end{center}
\caption{(a) top panel: Calculated time dependence of the displacement of an atom at the surface of ZnTe for a SAW at 1.25GHz propagating along [110]
and a chosen maximum displacement A=0.1nm. (a) bottom panels: corresponding calculated color map of the strain distribution. (b) Calculated $z$
profile of the strain distribution.} \label{FigSAW}
\end{figure}

Large in-plane dynamical strains at a precisely defined frequency can be generated by propagating
SAW and SAW device technology also offers simple approach for fabricating high quality factor
mechanical resonators. Following the  model presented in ref. \cite{Schuetz2015}, the strain
profile for a SAW propagating along the [110] direction of a zinc-blend material can be calculated
analytically. The strain profile for a SAW in the GHz range propagating in the surface of ZnTe
along [110] ($X$ direction) is presented in Figure~\ref{FigSAW}. Large oscillating in-plane strain
$\epsilon_{XX}$ oriented along the direction of propagation of the SAW can be obtained near the
surface of the sample. This strain field can be used to interact with a Cr spin located at less
than 100 nm from the surface. SAW also produce close to the surface a $Z$ component of the strain
that modifies the static magnetic anisotropy already present for a Cr in a CdTE/ZnTe QD.

To generate SAW on a Cr-doped QD, inter-digitated transducer (IDT) can be used to convert a
radio-frequency signal into a mechanical motion at a well defined frequency through the inverse
piezoelectric effect. The electromechanical coupling in an IDT depends on the square of the
piezoelectric coefficients $e_{ij}$. Compared to GaAs where SAW propagating along [110] direction
can be directly generated with an IDT, ZnTe has a very weak piezo-electric coefficient
$e_{14}\approx0.03Cm^{-2}$. A layer of a piezo-electric material, like hexagonal ZnO with its $c$
axis perpendicular to the sample surface, can be used to enhance the electromechanical coupling and
the performances of the transducer \cite{Thevenard2014}.

In the presence of an oscillating in-plane strain field along the [110] direction, $\epsilon_{xy}=\epsilon_{\perp}$, the non-diagonal coupling term
of the spin to strain Hamiltonian (\ref{epsilonperp}) for Cr becomes:

\begin{eqnarray}
{\cal H}_{Cr,\varepsilon\perp}=2c_4\varepsilon_{\perp}\frac{1}{2i}(S_+^2-S_-^2)
=-\frac{d_{\perp}}{2}\varepsilon_{\perp}(e^{i\frac{\pi}{2}}S_+^2+e^{-i\frac{\pi}{2}}S_-^2)
\end{eqnarray}

\noindent and the complete spin to strain Hamiltonian of Cr simplifies to

\begin{eqnarray}
{\cal H}_{Cr,\varepsilon}=(D_{0}+d_{\parallel}\varepsilon_{\parallel})S_z^2-\frac{d_{\perp}}{2}\varepsilon_{\perp}(e^{i\frac{\pi}{2}}S_+^2+e^{-i\frac{\pi}{2}}S_-^2)
\end{eqnarray}

\noindent where we defined the parallel $d_{\parallel}$ and perpendicular $d_{\perp}$ spin to strain susceptibility. A value of $d_{\perp}/2\approx 5
meV $ can be expected for a Cr atom in CdTe, at least two orders of magnitude larger than for NV centers in diamond where $d_{\perp}/2\approx 40 \mu
eV$ has been reported \cite{Barfuss2015}.

For small displacements, the strain is linear and we can quantize the perpendicular strain field $\varepsilon_{\perp}$. The corresponding strain
coupling Hamiltonian takes the general form:

\begin{eqnarray}
{\cal H}_{\epsilon\perp}=-\hbar\gamma_0^{\perp}(a+a^{\dag})(S_+^2+S_-^2)
\end{eqnarray}

\noindent where $\gamma_0^{\perp}$ is the transverse single phonon strain coupling strength and $a^{\dag}$ ($a$) the raising (lowering) operators for
phonons.

\begin{figure}[hbt]
\begin{center}
\includegraphics[width=3.5in]{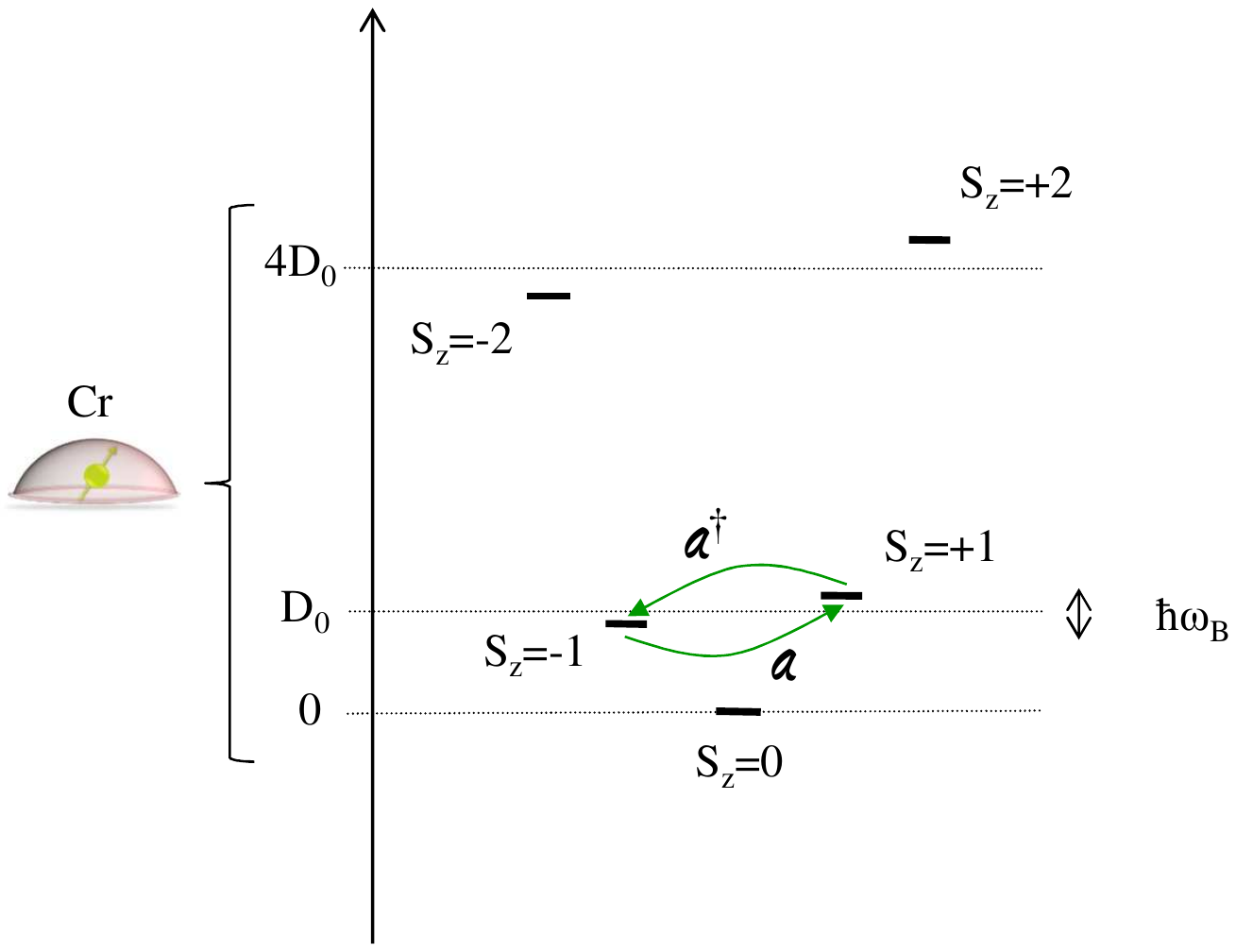}
\end{center}
\caption{Spin states of the Cr in a QD with large biaxial strain under a magnetic field applied along $[001]$. Local anisotropic in-plane strain produced by propagating SAW couples S$_z$=$\pm1$ states.}
\label{FigECr}
\end{figure}

In the situation where the coupling strength between the $qubit$ and the mechanical oscillator is
much smaller than the $qubit$ and oscillator frequency, it is a good approximation to drop the
counter rotating terms. This is the standard rotating wave approximation and the interacting
Hamiltonian takes the form:

\begin{eqnarray}
{\cal H}_{\epsilon\perp}=-\hbar\gamma_0^{\perp}(a S_+^2+a^{\dag}S_-^2)
\end{eqnarray}

For Cr in a strained QD where the spin states S$_z=0$, S$_z=\pm1$ and S$_z=\pm2$ are split by a
large magnetic anisotropy $D_0$, we can focus on the two-level spin subspace
$\{|+1\rangle,|-1\rangle\}$ only and assume that mechanically induced transitions to state
$|0\rangle$ or $|\pm2\rangle$ are not allowed due to the large zero field splitting (see
Figure~\ref{FigECr}). For a given spin, we write Pauli operators
$\sigma^{\pm}=|\pm1\rangle\langle\mp1|$ and $\sigma^z=|1\rangle\langle1|-|-1\rangle\langle-1|$.
Within this two-level subspace and within the rotating wave approximation, the interaction for a
single spin with the strain field takes on the generic Jaynes-Cummings form

\begin{eqnarray}
{\cal H}_{i}=\frac{\Delta_B}{2}\sigma^z+\hbar g_0(\sigma^+a+a^\dag\sigma-)+\hbar \omega_m a^\dag a
\end{eqnarray}
\noindent where a magnetic field parallel to the $z$ axis has been introduced to split the
$S_z=\pm1$ Cr spin levels by $\Delta_B=2g_{Cr}\mu_BB_z$ (with $g_{Cr}\approx2.0$). This magnetic
field allows to adjust the energy of the spin $qubit$ to the energy of a phonon of the SAW mode
$\hbar \omega_m$. Under this usual description common to many $qubit$ systems, the dynamics of the
${+1,-1}$ Cr spin $qubit$ can be obtained in closed-form.

For a coherent phonon field at frequency $\omega_m/2\pi$ (typically the one produced by a propagating SAW) tuned on resonance with the transition
between the $\pm1$ spin states, the spin to strain coupling can be rewritten

\begin{eqnarray}
{\cal H}_{Str,\perp}=\hbar\Omega_m cos(\omega_mt)(S_+^2+S_-^2)
\end{eqnarray}

\noindent Where $\hbar\Omega_m$, the Raby energy, describes the amplitude of the strain drive. Dynamical anisotropic in-plane strain provided by the
SAW \cite{Schuetz2015} could then be used for a direct coherent control of the $\{+1;-1\}$ spin $qubit$ under longitudinal magnetic field
\cite{Macquarrie2015}. Such mechanical coherent control could be probed with the resonant optical pumping technique presented in the previous
sections.

At high SAW excitation or in a SAW cavity, the large spin to strain coupling of Cr should permit to
obtain a Rabi energy of the order of magnitude of the $qubit$ energy. In this coupling regime,
called the strong driving regime, the rotating wave approximation is no longer valid and the
response of the two-level system to the driving field is the source of many interesting phenomena
which have not been studied in details until now.

SAW pulses (typically in the hundred ns range), could be used for a mechanical determination of the
dynamical spin to strain coupling and the coherence time of the ${+1,-1}$ spin $qubit$. This will
permit to determine the coupling regime that could be reached between a SAW cavity and a Cr spin.
In particular, the strong coupling regime could probably be reached for an ensemble of individual
Cr spins and the fundamental mode of a SAW cavity. For a detuned cavity and two distinct Cr-doped
QDs coupled to the same phonon mode of a SAW cavity, the mechanical motion can be used to induce a
long distance spin-spin coupling \cite{Bennett2013}.

\section{Conclusion}

We demonstrated that the spin of a Cr atom in a semiconductor can be probed and controlled
optically. In self-assembled QDs, the fine structure of the atom is dominated by a large magnetic
anisotropy induced by bi-axial strain in the plane of the QDs. Resonant optical excitation can be
used to control the spin of the atom. The Cr spin can be initialized by optical pumping, readout
through the resonant PL of the QD and the energy of any spin state can tuned by the optical Stark
effect. The spin relaxation of the Cr atom in the absence of optical excitation remains in the $\mu
s$ range at low temperature. The relaxation of a Cr spin in the exchange field of an exciton is
however much faster and dominated by spin flips induced by an interplay of the hole-Cr exchange
interaction and the interaction with acoustic phonons. The presence of these hole-Cr flip-flops
taking place in a few ns explains the efficient optical pumping observed under resonant excitation.

The S$_z=\pm1$ spin states of the Cr can form a spin $qubit$ with a large spin to strain coupling.
The possible optical control makes Cr a promising platform to study the interaction of the
${+1,-1}$ spin $qubit$ with SAW which are proposed as efficient quantum bus between different kinds
of $qubits$.

\section*{Acknowledgements}

The work was supported by the French ANR project MechaSpin (ANR-17-CE24-0024) and CNRS PICS contract No 7463. V.T. acknowledges support from EU Marie
Curie grant No 754303. The work in Tsukuba has been supported by the Grants-in-Aid for Scientific Research on Innovative Areas "Science of Hybrid
Quantum Systems" and for challenging Exploratory Research.

\end{document}